\documentclass[12pt,preprint]{aastex}

\begin{document}

\slugcomment{Astrophysical Journal}

\lefthead{Thompson}
\righthead{Properties of the NHDF}

\title{Star Formation History and Other Properties of the Northern HDF}

\author{Rodger I. Thompson}
\affil{Steward Observatory, University of Arizona,
    Tucson, AZ 85721}

\begin{abstract}

The original analysis of the star formation history in the NICMOS Deep
images of the NHDF is extended to the entire NHDF utilizing NICMOS and
WFPC2 archival data.  The roughly constant star formation rate from
redshifts 1 to 6 found in this study is consistent with the original
results. Star formation rates from this study, Lyman break galaxies and
sub-mm observations are now in concordance  The spike of star formation
at redshift 2 due to 2 ULIRGs in the small Deep NICMOS field is
smoothed out in the larger area results presented here.  The larger
source base of this study allows comparison with predictions from
hierarchical galaxy formation models.  In general the observation are
consistent with the predictions.   The observed luminosity functions at
redshifts 1-6 are presented for future comparisons with theoretical
galaxy evolution calculations. Mid and far infrared properties of the
sources are also calculated and compared with observations.  A
candidate for the VLA source VLA 3651+1221 is discussed.

\end{abstract}


\section{Introduction} \label{sec-in}

The Hubble Deep Fields (HDFs) are rich sources of data on cosmology
and the evolution of galaxies.  The large number of observations at
many different wavelengths make the Northern HDF (NHDF) particularly
useful. The NHDF is the only HDF that has complete coverage with both
WFPC2 and NICMOS.  The NICMOS coverage, however, is not to the same
depth as the WFPC2 due to the smaller area of the NICMOS Camera~3. This is 
somewhat compensated for by the red nature of
evolved galaxies and the redshift of visible light into the infrared
bands for young blue galaxies.

This work studies the star formation history in the NHDF and is an
extension of a similar study utilizing the smaller NICMOS Deep field
\citep{thm01} (hereinafter TWS).  The increased area and number of
galaxies greatly improves the statistical significance of the results
over TWS.  Shallower coverage, however, increases the photometric
error, making it comparable or dominant over the large scale structure
error which decreased due to the larger field.

Other researchers, \citet{lanz96,fs99,lanz02} have utilized the same
data set, plus additional ground based data to investigate star
formation history in the NHDF. Our analysis differs significantly from
those studies by including extinction in the SED templates used to
determine the photometric redshifts.  This produces results quite
different from \citet{lanz02} who do not include extinction.  Contrary
to \citet{lanz02} we find that the star formation rate (SFR) is
essentially constant in the epoch from z = 1 to z = 6 and that this
rate is significantly larger than the present day SFR \citep{lil96},
consistent with the results presented in TWS.

\section{Observations} \label{sec-obs}

Observations utilized in this study are all from the Hubble Space
Telescope (HST) archives.  The first set is the processed WFPC2 images
of the NHDF produced by \citet{wil96}.  The second set is the archival
data from the NICMOS survey of the entire NHDF by \citet{dic00}.  The
NICMOS F110W and F160W data were reprocessed as described in
\S~\ref{sec-dr}.  Only the areas corresponding to the three wide field
WFPC2 chips are used since the WFPC2 PC chip images have substantially
different signal to noise statistics.  NICMOS Deep NHDF images and data
reductions from \citet{thm99} and TWS are used in the error analysis.

All NICMOS images were taken in the SPARSE64 mode with 24
samples after the first read and integration times of 1344
seconds.  Images at 8 positions completely covered the NHDF.
At each position 9 images were taken in a dithered pattern.  The 8
positions with 9 dither points in the F110W and F160W filters
required 144 exposures.

\section{Data Reduction} \label{sec-dr}

The data reduction procedures are almost identical to the procedures
described in \citet{thm99} and TWS.  For this reason they will only be
briefly outlined except where there are differences.  The WFPC2 images
were not altered from the Version 2 images available in the HST
archives.  The NICMOS images, however, were reduced from the raw images
obtained in the HST cycle 7 program 7817 by Mark Dickinson.  The images
were examined for effects of cosmic ray persistence from SAA passages
and 17 of the 144 images were rejected.  The background flux was
removed by taking the median of the images in each filter that were not
affected by cosmic ray persistence and subtracting it from each image.

The 9 dithered images for each of the 8 positions were combined using
the standard Drizzle procedures with 0.1 arc second pixels, one half
the linear size of the NICMOS Camera~3 pixels. The IDL procedure IDP3
\citep{lyt99} produced a mosaic of the 8 images. IDP3 also reduced the
WFPC images to the NICMOS resolution with a bi-cubic spline
interpolation and aligned them with the NICMOS mosaic. Inspection of
the images showed regions of increased noise at the boundaries between
images.  These areas were masked out as shown in Figure~\ref{fig-mask}.
Since this study is primarily a statistical study, the small amount of
masked area does not affect the conclusions.

\includegraphics{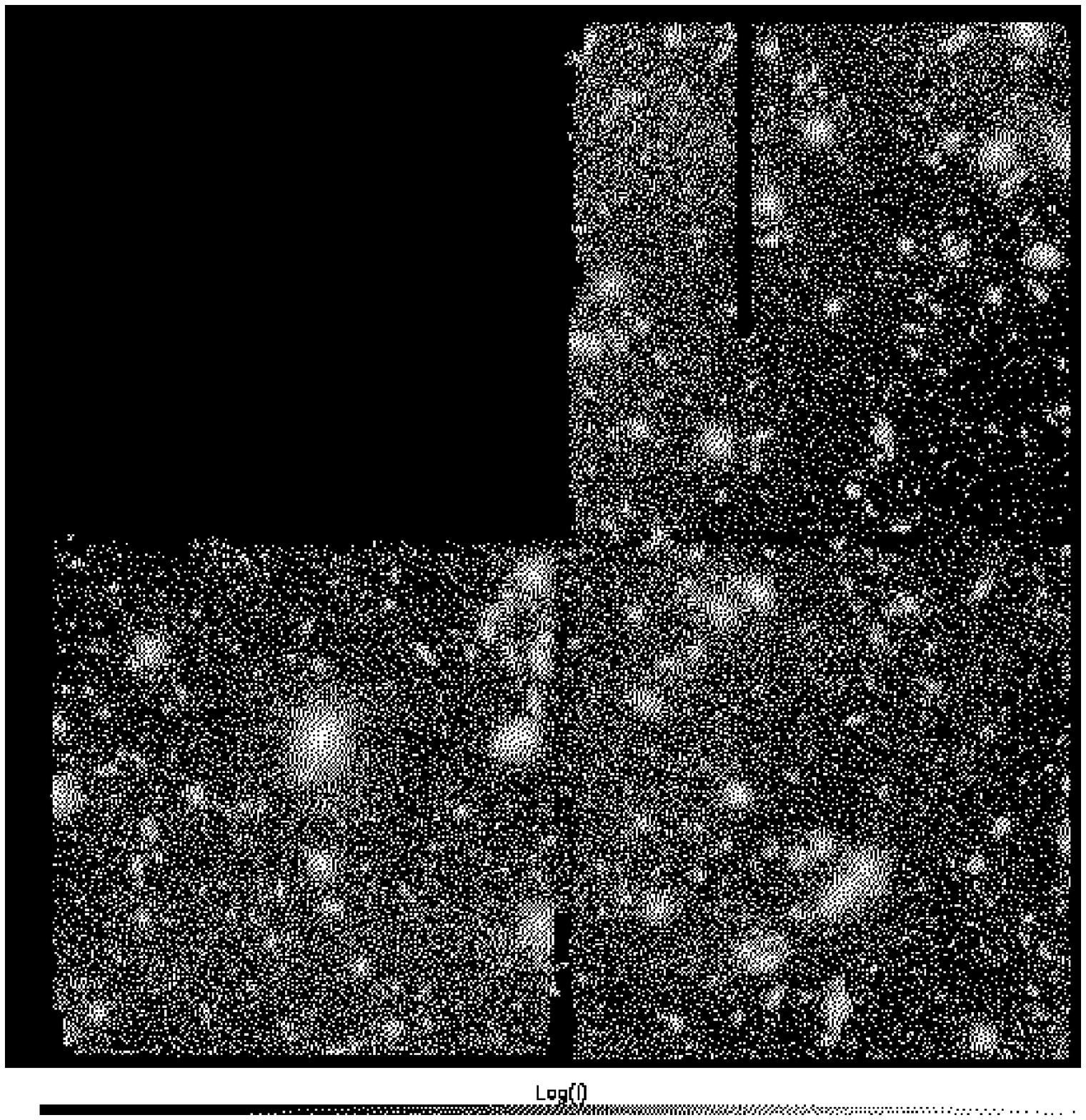}

\begin{figure}

\caption{The masked areas shown on the WFPC2 F814W filter image of 
the NHDF.}

\label{fig-mask}
\end{figure}

\clearpage

Source positions were found using the method of \citet{sza99} as
described in TWS.  A Gaussian fit to the histogram of pixel values
determined the global variance of the image mosaic in each filter.  The
histogram is dominated by the background noise with true sources
creating a deviation from the Gaussian fit at large positive values. The
1 $\sigma$ noise levels in both the F110W and F160W NICMOS filters are
$5 \times 10^{-4}$ ADUs per second.  For comparison the NICMOS Deep
Field has a 1 $\sigma$ noise level of $2.2 \times 10^{-4}$ ADUs per
second.  $5 \times 10^{-4}$ ADUs per second corresponds to $1.4 \times
10^{-9}$ Janskys.  The drizzling procedure, however, introduces
correlation,  therefore, the true noise level is about 1.6 times larger
than the measured Gaussian noise.

The Szalay source identification procedure requires images with a local
variance of one.  The local variances for the F110W and F160W fields
were found by drizzling and mosaicing the flat fields of the two
filters in exactly the same way as the images. This accounts for
differences in sensitivity over the area of the NICMOS Camera 3
detector. The flats have a median of 1, therefore, the variance images
also have a median of 1.  The ratio of the imaged fields to the
variance image was scaled to the average variance creating a local
variance of 1. The variance in the WFPC filters was assumed constant
across the field. This procedure differs slightly from the procedure
used in TWS.

\begin{figure}

\plotone{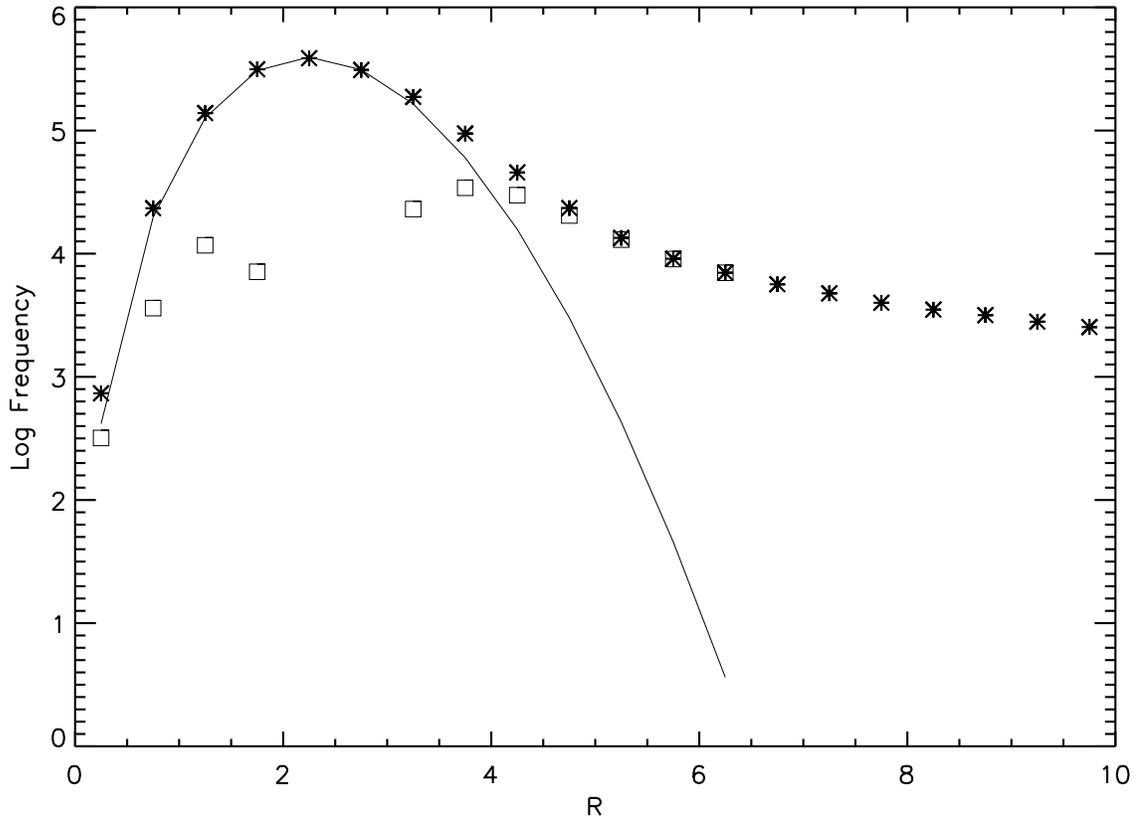}

\caption{R image pixel distribution from the Szalay procedure. The
asterisks represent the actual data and the solid line is the expected
$\chi^2$ distribution with six degrees of freedom.  The boxes show the
difference between the data minus the expected distribution.  The long
tail at high R is the real source distribution.}

\label{fig-r}
\end{figure}

The threshold value of the Szalay R parameter, $\sqrt{\chi}$, was set
to 4.5, higher than the 2.3 value used in TWS but more appropriate for
the six dimensional $\chi^2$ analysis of the combined WFPC and NICMOS
fluxes. Unlike TWS all six fluxes are included in the Szalay procedure
to insure that small blue galaxies with little infrared flux are not
missed in the source extraction.  The source selection criterion for
determining the star formation history (\S~\ref{ssec-ssc}), however,
does discriminate against galaxies with no F110W or F160W flux.
Figure~\ref{fig-r} shows the distribution of R image pixels.  All
pixels with R values higher than or equal to 4.5 are considered
legitimate source pixels.  Pixels with R values less than 4.5 are
assigned random values with average values $10^6$ below the valid
pixels.  This forces the source extraction program to only consider
pixels chosen by the Szalay procedure.  The source extraction program
crashes if all non-valid pixels are set to zero.

The source extraction program SExtractor (SE) \citep{ber96} extracts
the sources in two passes. The first pass on the R has the extraction
threshold  set at 4.0, selecting  all valid pixels.  As in TWS the
DETECT MINAREA parameter is set to 3 so a true source must have 3
contiguous valid pixels.  The source positions, except for the
contiguous area criterion are, therefore, determined by the Szalay
procedure, \emph{not by SE}. SE then runs again on the six photometric
images in a mode that extracts sources in the pixel pattern determined
by the first pass. Run in this manner, SE does not produce daughter
objects. Sources which have several components may be broken into
individual objects by this process.  Although listed as separate
objects in the catalog the star formation history is unaffected as long
as the calculated photometric redshift is the same for both components.
In some cases a separate listing for two components is an advantage,
allowing different extinctions for different components.
\S~\ref{ssec-ssc} discusses additional criteria for final source
selection.

The PHOT-APERTURES parameter in the SE configuration file was set to 6,
10, 15 to produce 0.6$\arcsec$, 1.0$\arcsec$, and 1.5$\arcsec$ diameter
aperture fluxes.  The photometric, redshift, extinction and SED
calculations use the 0.6$\arcsec$ aperture fluxes. The total flux for
determining the SFR is found by summing all of the pixels identified by
the Szalay procedure as belonging to a single source.

\section{Photometric Redshift, Extinction and SED Determination} 
\label{s-pres}

The basic procedure for photometric redshift, extinction and SED
determination is a $\chi^2$ comparison of the observed fluxes in the
six bands to fluxes calculated from numerically redshifted and
extincted SEDs.  As described in TWS we draw our templates from three
sources.  The first source is the four observed SEDs of \cite{cww80}
utilized by several authors.  These galaxy SEDs have been corrected for
galactic extinction but not for extinction in the galaxies themselves.
The galaxies, however, were selected on the basis of their very low
internal extinction. The unreddened SED of the set of mean SEDs of
\cite{cal94}  provides an additional observed active star-forming
galaxy template (\cite{cal99x}).  A final and even hotter template is a
50 million year old continuous star formation SED calculated from the
Bruzual and Charlot models (\cite{bc96}) with a Salpeter IMF and solar
metallicity. This theoretical SED does not have emission lines,
therefore we have added  H$\alpha$, (O[III]+ H$\beta$) and O[II]
emission lines by scaling up the lines from the Calzetti SED by the
ratio of the UV fluxes in the Calzetti and Bruzual--Charlot SEDs.
Template 6 is substantially bluer than the Calzetti SED.
  
Extinction is calculated from the starburst and star forming galaxy 
optimized obscuration law of \citet{cal94}.  Since the primary purpose
of this paper is a determination of the star formation rate it is
appropriate to utilize an extinction law determined from star
forming galaxies.  The absorption due to extragalactic neutral hydrogen
is calculated from an updated version of the formulation of
\citet{mad96}.  The calculated fluxes are then interpolated between the
6 SEDs to produce 51 different SED fluxes for each redshift and
extinction point. The details of this procedure are given in TWS. The
grid for the $\chi^2$ analysis includes 100 different redshifts between
0 and 8, 15 different extinctions between E(B-V) of 0 to 1.0 and 51
different interpolated template types for a total of $7.65 \times 10^4$
choices.  The resulting redshifts, extinctions and SEDs are given in
Table~\ref{tab1}.  The earliest SED has a value of 1.0 incremented by
0.1 to the latest SED of 6.0.  Integer values refer to the 6 basic SED
templates.  Extinctions in E(B-V) are incremented by 0.02 from 0 to 0.1
and by .1 from 0.1 to 1.0, the maximum used in the analysis.

The $\chi^2$ procedure is identical to TWS except that negative fluxes
are not replaced by zero.  The technique minimizes the $\chi^2$
residuals between the observed fluxes and those predicted by the
numerically redshifted, extincted and Lyman absorption attenuated SEDs.
It alters the usual error term in the denominator by adding a second
term proportional to the measured flux.   The $\chi^2$ residual is

\begin{equation} \chi(z,E)^{2} = \sum_{i=1}^{6} \left(\frac{(f_i - A 
\cdot 
fmod(z,E)_i)} {\sqrt{\sigma_i^2 + (0.1f_i)^2}}\right)^2 \label{eq:chisq} 
\end{equation}

\noindent The index~i refers to the six fluxes, $f_i$ is the measured
flux and $fmod(z,E)$ is the flux predicted by a template at a redshift
of z and extinction E(B-V) = E.  Since this is not a formal $\chi^2$
calculation the quantitative probabilities associated with $\chi^2$
values are not strictly valid. The normalization constant A minimizes
$\chi(z,E)^{2}$.

\begin{equation} A =  \sum_{i=1}^{6} \frac{f_i \cdot fmod(z,E)_i} 
{\sigma_i^2 
+ (0.1f_i)^2} /\sum_{i=1}^{6} \frac{(fmod(z,E)_i)^2}
	{\sigma_i^2 + (0.1f_i)^2} \label{eq:a} \end{equation}

The limit of the expression at very low flux levels is the standard
form with the formal background, $\sigma$, dominating the denominator
and at high flux levels it is the flux difference between the
observations and the model divided by 10$\%$ of the flux instead of the
$\sigma$.  At high flux levels the errors in the fit are proportional
to flux since the dominate errors are intra-pixel sensitivity
variations and systematic flux errors. The accuracy of the NICMOS flux
levels may be better than this. It is estimated to be 4 - 5$\%$ for
this level of dithering but the output does not appear to strongly
depend on the precise coefficient of the flux in the denominator of
equation(~\ref{eq:chisq}).

\subsection{Photometric Redshift Accuracy} \label{ss-prac}

Photometric redshifts have gained a reasonable reputation for accuracy,
particularly at high redshifts where the clear signature of the Lyman
break is easy to observe.  Two tests of the photometric redshifts
derived in this work are comparison to available spectroscopic NHDF
redshifts and comparison with previous NHDF photometric redshifts (TWS,
\citet{fs99}, \citet{wng98}).

\subsubsection{Comparison with Spectroscopic Redshifts} \label{ss-csr}

\begin{figure}

\plotone{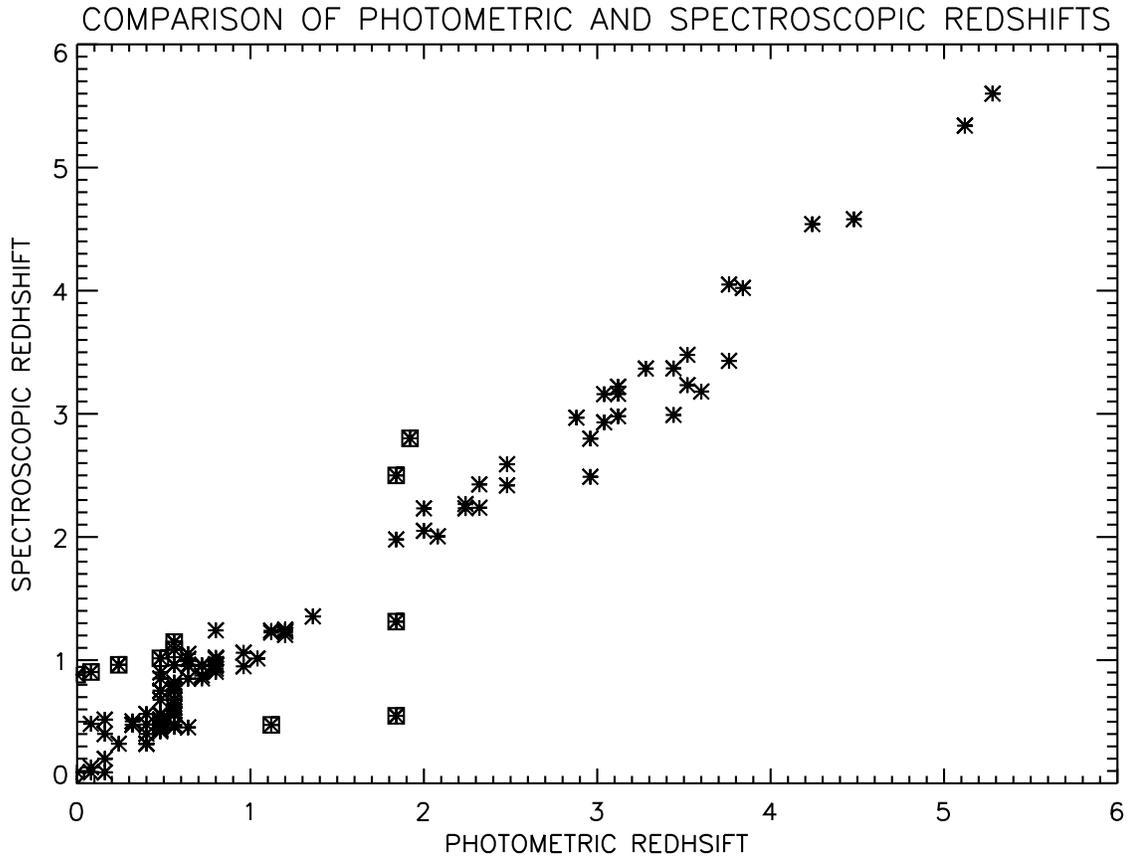}

\caption{Plot of the spectroscopic redshifts from \citet{chn00} versus
the photometric redshifts determined in this work. Some high redshift
objects have been added from other sources. Objects that have redshift
differences greater than 0.5 are overplotted with a square symbol and
are discussed in the text and Table~\ref{tab-err}.}

\label{fig-zcomp}
\end{figure}

The primary source of spectroscopic redshifts is \citet{chn00},
enhanced and modified by \citet{chn01}. The small reductions in the
field area described in \S~\ref{sec-obs} reduces the number of
comparison redshifts to 135.  Figure~\ref{fig-zcomp} shows the results
of the comparison.  The standard deviation is 0.29 in redshift,
however, it is not constant with redshift.  Most of the error occurs in
the redshift range between 0 and 2, as is expected, since the prominent
Lyman break does not appear in our data for that range.  Note that
galaxies in the redshift range between 0 and 0.5 are not used in this
study.

Catastrophic redshift errors, errors of more than 0.5 in redshift, are
overploted with squares in Figure~\ref{fig-zcomp}. Analysis of the 11
catastrophic failures reveals three basic types of failures.  The
first type is where there is no clear minimum in the $\chi^2$ value
along the range of redshifts, extinctions or SED templates.  These are
characterized by a steadily decreasing $\chi^2$ value with a parameter,
such as template, until the end of the parameter space is reached.  This
failure of the parameters to span the space of galaxy types is seen
in 6 galaxies. The second category contains 5 galaxies that are
most probably superpositions of two galaxies that have not been
separated by the SE source extraction process.  Both galaxies are
contained in the LRIS 1$\arcsec$ slit width used in the spectroscopic
observations \citep{chn00}.  In this case the spectroscopic redshift
would be determined by the galaxy with the most prominent spectral
features while the photometric redshift would be most influenced by the
galaxy with the highest continuum level.

The third failure in 2 galaxies is an apparent mild degeneracy in
parameters which tends to favor a redshift of about 0.5 for galaxies
with spectroscopic redshifts between 0.5 and 1.0.  These galaxies
appear to be well fit at the correct redshift with an intermediate SED
and low value of extinction or at a redshift of $\sim$.5 with a very
hot template and a high value of extinction.  This degeneracy creates a
pile up of galaxies at redshifts of 0.48 and 0.56. Examination of the
photometric redshifts for these sources where the extinction is forced
to be zero did not give statistically better fits except in the extreme
cases such as those in Table~\ref{tab-err}.  This type of error can
reduce the calculated SFR in the redshift 0.5 to 1.5 bin by moving some
sources out of the bin to a redshift of 0.48 or reducing the calculated
UV flux for a source that has been moved to a lower redshift
erroneously.  This would make the SFR for redshift a  lower limit,
subject to other errors. Note that some galaxies have multiple failure
modes so that the number of failures listed is larger than the number
of failed galaxies.

Table~\ref{tab-err} lists the 11 galaxies with redshift errors larger
than 0.5 along with their parameters and the probable cause of the
error.  The general conclusion is that the photometric redshifts are quite
adequate for statistical studies but can be subject to
significant errors for individual galaxies.

\subsubsection{Comparison with Previous Photometric Redshifts}

The comparison between the spectroscopic and photometric redshifts is
dominated by low redshift fairly bright galaxies except for the few
faint, high redshift galaxies that have spectroscopic redshifts through
very long spectroscopic observations on the largest ground based
telescopes.  It might be expected that the agreement on bright sources
may be better than for the more numerous faint galaxies.  One test is a
comparison between the photometric redshifts obtained in the Deep
NICMOS field (TWS) with those found in this study.  The NICMOS F110W
and F160W images are completely independent between the two data sets
while the WFPC2 image are in common except that KFOCAS was used for
source extraction in TWS and SE is used here.  The comparison primarily
determines the differences introduced by noise in the F110W and F160W
images.  Redshifts from the NICMOS deep NHDF images should be more
accurate than the redshifts from the present study due to the higher
signal to noise of the deep images.

\begin{figure}

\plotone{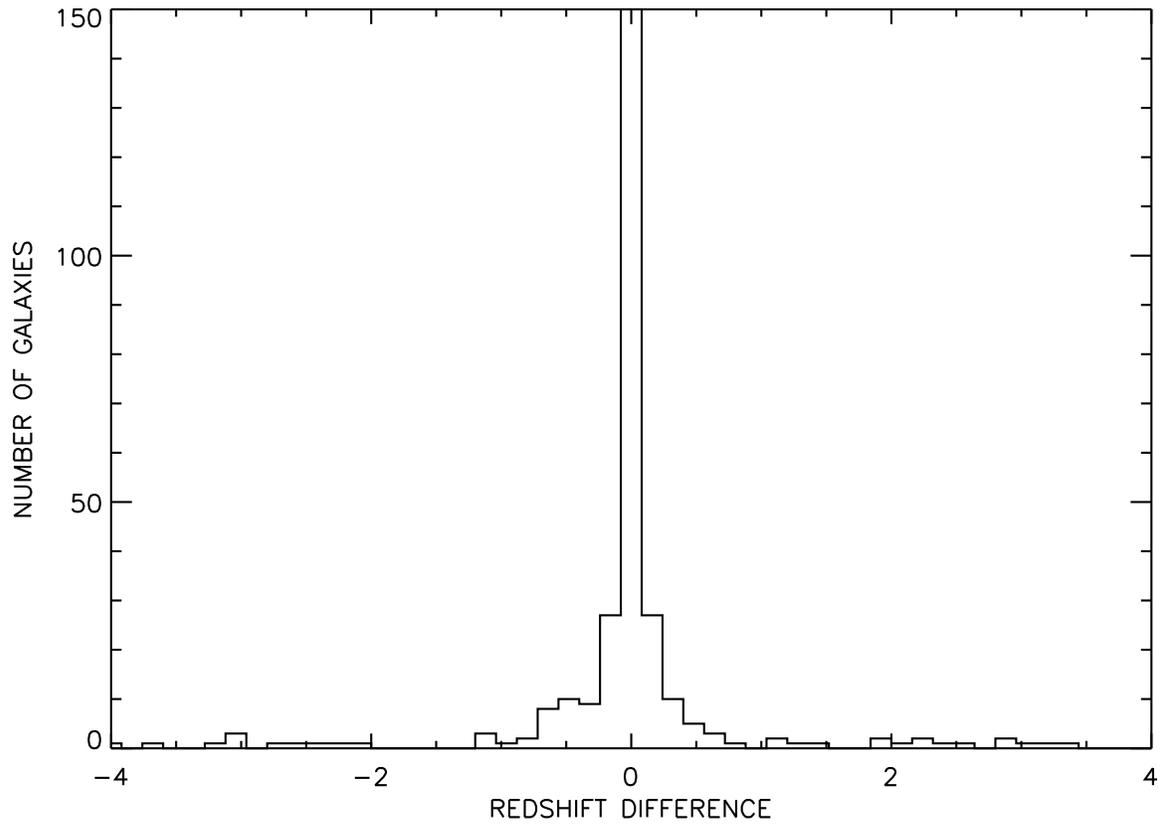}

\caption{A histogram of the values of the Deep NICMOS NHDF photometric
redshifts minus the photometric redshifts determined from this study.}

\label{fig-zhist}
\end{figure}

A total of 283 galaxies in the Deep images also appear as legitimate
galaxies in this set of images. Figure~\ref{fig-zhist} shows a
histogram of the differences in the redshifts between the deep and
present images.  The differences are peaked at zero as expected but do
not appear to be a Gaussian distribution away from the peak.  The
differences are reasonably equally distributed between positive and
negative values with a slight skew toward negative values (the present
redshift higher than the deep redshift).  The percentage of
"catastrophic failures", a difference greater than 0.5, is $17\%$.
Some of the failures create large differences in redshift.  Examination of
some cases with large differences reveals significant differences in
either the F110W or F160W flux. As expected there is little variation
in the WFPC2 fluxes measured by KFOCAS and SExtractor. The $17\%$
significant redshift error from this comparison is higher than the
$8\%$ error found from the spectroscopic redshift comparison.  In the
final error analysis we will assume that $17\%$ of the galaxies have
redshift errors that are randomly distributed up to a maximum of 4.

\subsubsection{Comparison with Photometric Redshifts from Other Work}

\begin{figure}

\plotone{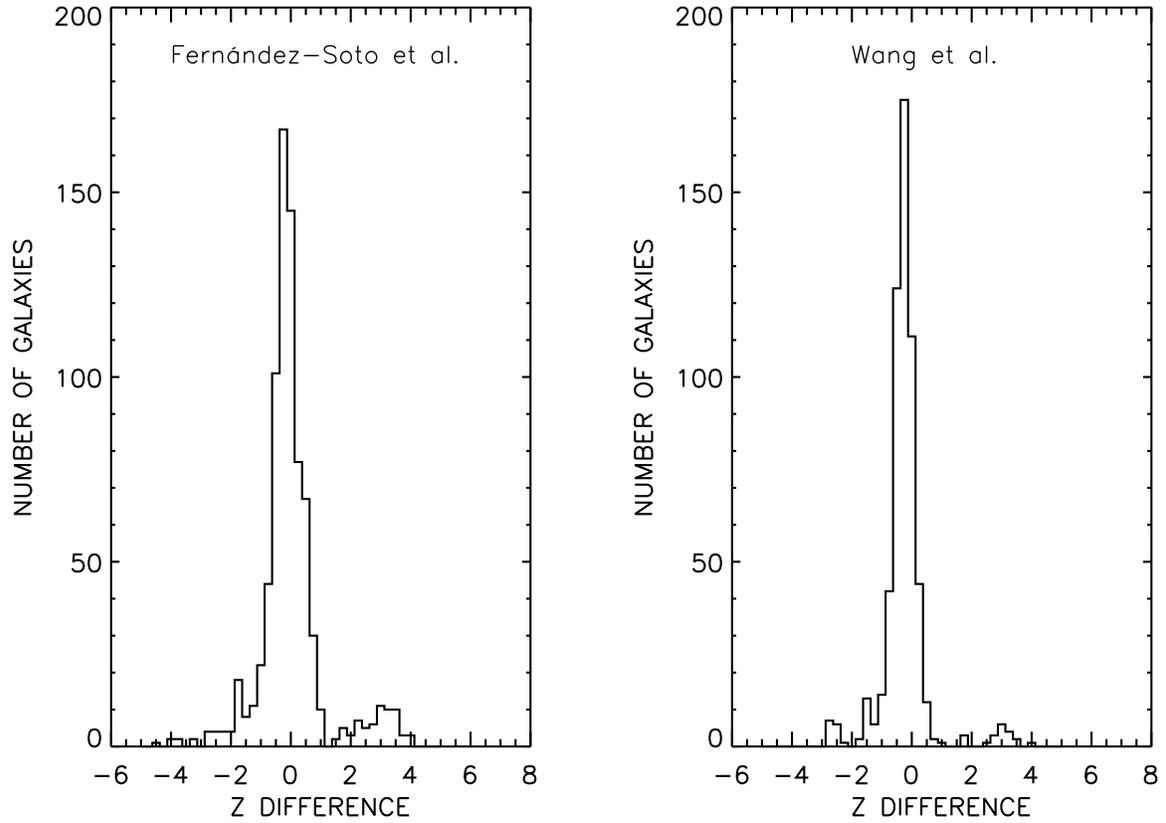}

\caption{Histograms of the values of the photometric redshifts
determined from this study minus the redshifts from \citet{fs99} on the
left and \citet{wng98} on the right. In both case the redshifts from
the cited works were subtracted from the redshifts found in this work.}

\label{fig-zlwcomp}
\end{figure}

There have been several other determinations of photometric redshift in
the NHDF but only two \citep{wng98,fs99} have readily accessible
catalogs of redshift and object position.  Figure~\ref{fig-zlwcomp}
shows the comparison between this work and the redshifts obtained by
the two groups.  Coincidence of objects was determined by position
correspondence within 0.3$\arcsec$.  The redshifts of both previous
studies are offset to larger redshifts than this study.  This effect is
most probably due to both groups not correcting for extinction.
Without extinction the only mechanism to make a galaxy redder is to
redshift it or choose an earlier template.

\clearpage

\section{Catalog of Individual Source Properties} \label{sec-cat}

Table~\ref{tab1} lists the individual sources, ordered by RA, that
satisfy the criteria listed in \S\ref{ssec-ssc}.  Column 1 gives the
NICMOS identification number. Column 2 is the WFPC2 identification
number of the galaxy from \cite{wil96} if there is positional
coincidence within $0.3\arcsec$. Columns 3 and 4 are the the redshift
and extinction. Column 5 contains the SFR in solar masses per year as
determined in \S~\ref{sec-dsfr}. The bolometric luminosity of the
galaxy is in column 6. The flux of a galaxy is obtained by integrating
over the unextincted selected template scaled by the factor A from
equation(~\ref{eq:a}).  The bolometric luminosity then follows from the
redshift and our adopted cosmology. The fraction of the luminosity that
is extincted and therefore goes into far infrared flux is in column 7
followed by the calculated 6, 15 and 850 $\micron$ flux (mJy) in
columns 8, 9 and 10 for comparison with measured ISO and SCUBA fluxes
(\S~\ref{sec-ir}).  The luminosities, ISO fluxes and SCUBA fluxes of
galaxies with a redshift of zero are all set to zero.

Columns 11 and 12 give the template number T of the best fit, and the
modified $\chi^2$ value of the fit from Equation~\ref{eq:chisq}. It
should be noted that the distribution of the modified $\chi^2$ values
will not rigorously follow a true $\chi^2$ distribution. The values are
provided to give a qualitative indication of the relative goodness of
fit for the best fit values of redshift, template type and E(B-V).
Column 13 has the total F160W AB magnitude, determined by adding all of
the pixels designated by SE as part of an object, followed by the 0.6
aperture F160W magnitude (Ap. mag) in column 14. Note that variations
in local background as interpreted by SE can result in an aperture
magnitude brighter than a total magnitude for faint galaxies.  If no
F160W magnitude is listed the object had a zero or negative measured
F160W flux. Columns 15 and 16 are the right ascension and declination
positions of the object. The RA listing contains only seconds and the
DEC listing only minutes and seconds. 12$\fh$ 36$\fm$ should be added
to the RA and 62$\fdg$ to the DEC. Note that a few sources have RA
values of 12$\fh$ 37$\fm$ plus seconds.  These sources are found at the
end of the catalog with seconds values between 0 and 1.2.

\subsection{Source Selection Criteria} \label{ssec-ssc}

Data reduction is performed on all  sources detected by the Szalay
procedure. Additional criteria for source selection are introduced
before sources appear in the catalog of objects used for determining
the star formation history.  First all sources that are known stars are
rejected.  Next all objects that are too near the edge of the image to
insure that no flux is lost are rejected.  This step uses the SE
internal flags and visual inspection of the image.  Next a list of
suspect objects are rejected from a visual inspection of the images.
These include objects with significant overlap that have been counted
as a single object by SE and stellar diffraction spikes that were
counted as objects.  Next all objects with SE internal flags  of 16 or
greater, indicating that the photometric aperture was corrupted by some
means, are rejected.  Next is a criterion that all valid sources must
have a signal to noise of 3.5 or greater in one band or a signal to
noise of 2.5 or greater in two bands.  Finally an accepted source is
required to have a flux greater than $5 \times 10^{-4}$ ADUs per second
in both the F110W and F160W bands.  All sources that meet these
criteria are included in the catalog but only sources with redshifts
between 0.5 and 6.5 are used in the star formation history analysis.

Two other comments are in order on source selection.  At this point, for
the small number of objects that have known redshifts, the photometric
redshift is replaced by the spectroscopic redshift in the analysis
procedures.   Second the bright galaxy NHDF 3-610 is rejected from the
analysis due to impingment of the diffraction spike from the adjacent
star and a clear overlap with another galaxy.  For this reason, even
though it is quite bright, it does not appear in Table~\ref{tab1}.

\subsection{Distribution of Source Properties}

\begin{figure}

\plotone{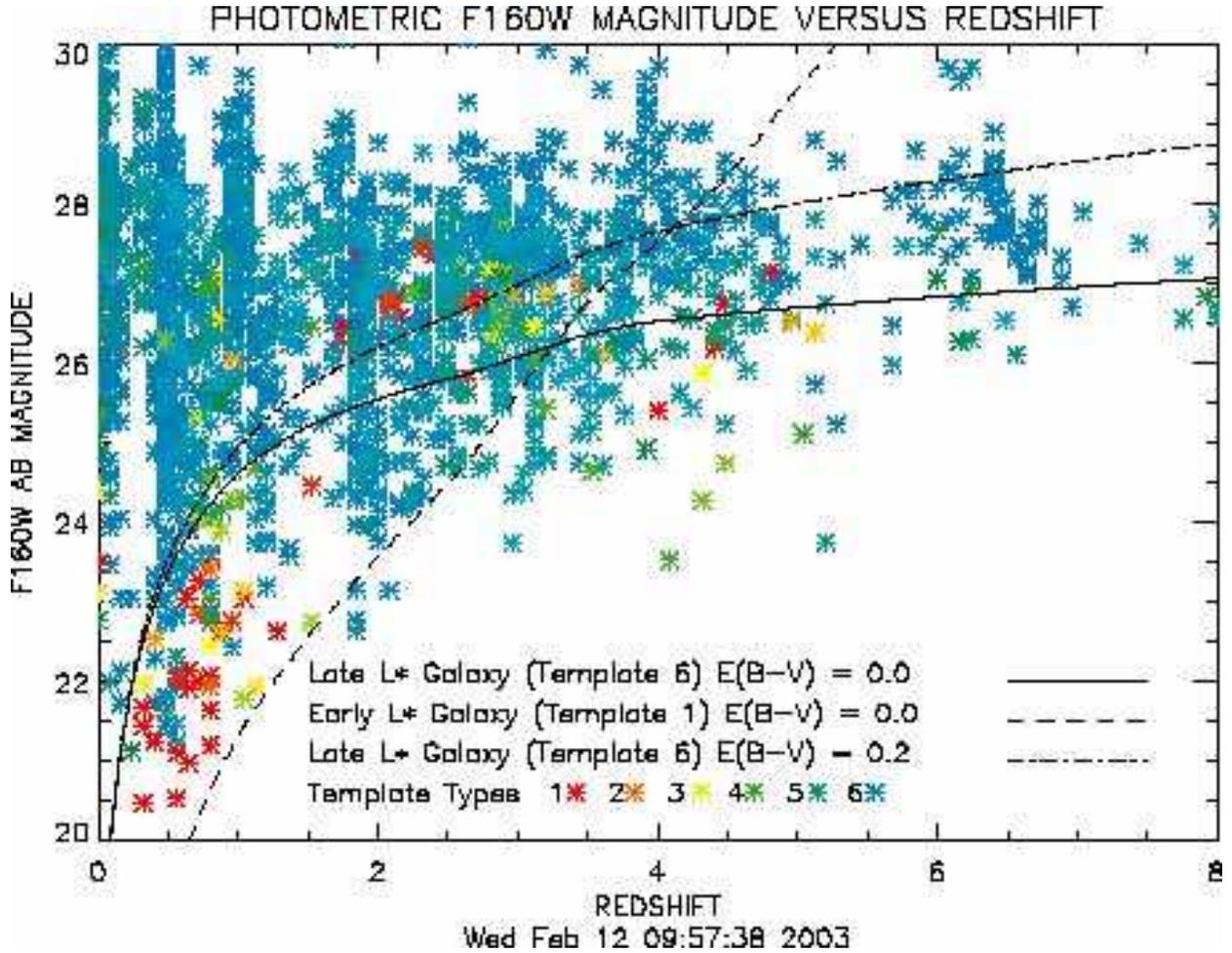}

\caption{The distribution of F160W AB 0.6$\arcsec$ diameter aperture
magnitudes versus photometric redshift. The solid and dashed lines
indicate the F160W aperture magnitude of an early and late-type L$^*$
galaxy. The dash dot line shows the track of a late-type L$^*$ galaxy
with an extinction of E(B-V) equal to 0.2.  L$^*$ is defined as a total
luminosity of $3.4 \times 10^{10}$ L$_{\sun}$}

\label{fig-hmag}
\end{figure}

Figure~\ref{fig-hmag} shows the distribution of F160W magnitudes with
redshift and the calculated tracks of an early, a late, and an
extincted late L$^*$ galaxy.  The early galaxy is template 1.0, the
late galaxy template 6.0 and the extincted late galaxy is template 6.0
with an extinction of E(B-V) = 0.2.  The sources are color coded with
the earliest type galaxy deep red and the latest galaxy deep blue as
indicated by the legend on the figure.  As expected the majority of the
early type galaxies lie at low redshift with some intermediate type
galaxies up to redshifts of five.  The majority of galaxies have
template numbers between 5 and 6.

Particularly at low values, there is apparent redshift banding.  The
band at zero redshift is due to galaxies with $\chi^2$ values that were
still falling as they approached zero redshift, indicating that the
were not matched by any of the galaxy templates.  The band at redshift
0.5 does correspond to an over density observed in the spectroscopic
redshift analysis of \citet{chn01} which also sees a small over density
at a redshift of 1. It is tempting to interpret the banding in
Figure~\ref{fig-hmag} as sheets or nodes in large scale structure
encountered by the pencil beam of the NHDF but the redshift analysis in
\S~\ref{s-pres} does not seem to warrant that interpretation. This is
particularly true for the band at 0.5 which may be due to the effect
discussed in \S~\ref{ss-csr}. Investigation of the significance of the
banding will be deferred to a later paper.
  
\begin{figure}

\plotone{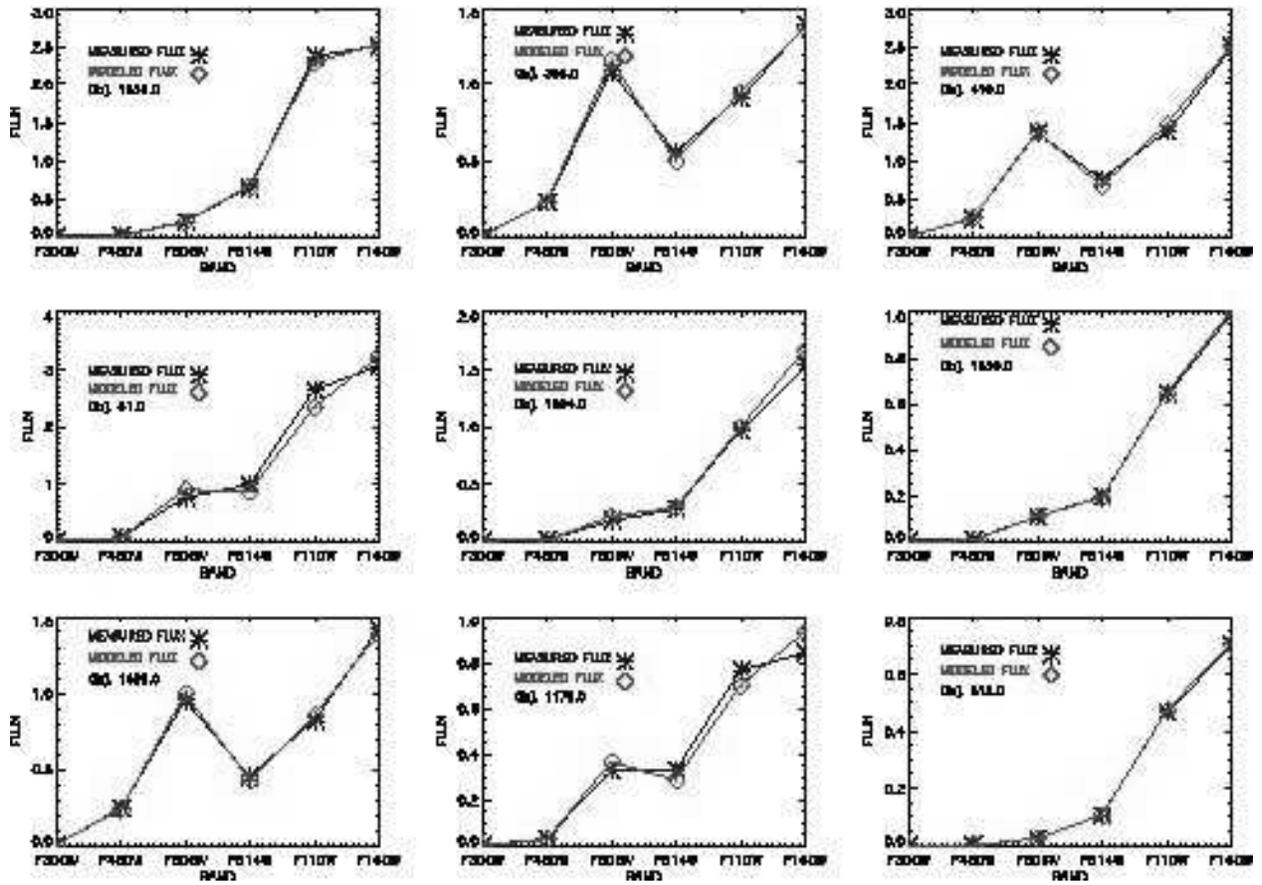}

\caption{The best fit fluxes matched to the observed fluxes for 9
galaxies that are significantly brighter than the L* galaxy line.  
Luminosities are given in L$_{\sun}$.}

\label{fig-fits}
\end{figure}

Most of the galaxies in Figure~\ref{fig-hmag} fall above the lines of
the L$^*$ ($3.4 \times 10^{10}$ L$_{\sun}$) galaxy tracks.  Past a
redshift of $\sim$3 there are several galaxies that lie significantly
below the L$^*$ galaxy line.  This is expected since there certainly
are galaxies brighter the L$^*$.  It is legitimate to ask whether these
galaxies have non-physical parameters and whether their position in the
diagram is due to a bad fit with the models.  Fig.~\ref{fig-fits} shows
the fit between the observed fluxes of the 9 galaxies that most exceed
L$^*$ and the best fit fluxes from the photometric redshift program.
In all cases the fit is excellent, therefore, their position in the
diagram is not due to a bad fit.  Table~\ref{tab-lstar} gives a summary
of the properties of the nine sources.  Although all but one galaxy has
the luminosity of a Luminous InfraRed Galaxy (LIRG), the low extinction
does not re-emit a large enough fraction of the luminosity to put them
in that category.  All but 2 galaxies fall in the top 50 most luminous
galaxies in the field but none of them are in the top 10.  This leads
to the conclusion that the physical properties of the galaxies are
realistic and that there is no reason to exclude them as legitimate
sources based on non-physical parameters.

Note that 398.0 and 410.0 are probably two parts of the same galaxy.
Their photometric redshifts of 3.04 and 2.96 are similar but not
exactly the same. The spectroscopic redshift for the the galaxy is
2.799, close to the photometric redshift. The spectroscopic redshift is
the redshift used in calculating the luminosity for source 410. If the
spectroscopic redshift for source 410 is also the proper redshift for
source 398, the 398 luminosity should be reduced by a factor of 0.74.

There are three sources without equivalent WFPC2 identifications.
Source 398 is an extension of source 410 identified as WFPC2 4-555.1.
The closest WFPC2 source to 398 is actually 4-555.11 which is
0.32$\arcsec$ away, just beyond our association radius of
0.3$\arcsec$.  Similarly the closest WFPC sources for 1630 and 1636 are
3-367.0 (0.33$\arcsec$) and 3-839 (0.40$\arcsec$).  Although not listed
in Table~\ref{tab1},the identifications are listed in
Table~\ref{tab-lstar} since inspection of the images clearly indicates
that they are the proper associations.  It is often the case that the
F160W center of a galaxy is offset from the optical position.  The
F160W position is more representative of the underlying stellar
distribution while the optical position measures rest frame UV star
formation activity.

\section{Determination of the Star Formation Rates} \label{sec-dsfr}

As in TWS the SFR is determined from the broad band 1500 $\AA$ flux
relation given by \citet{mad98}.

\begin{equation} UV_{1500} = 8.0 \times 10^{27} \cdot SFR(M_\odot /yr)  
\ 
\mathrm{ergs\ second^{-1}\ Hz^{-1}} \label{eq:sfr}
\end{equation}

\noindent The value of the UV flux is determined from the 1500 $\AA$
flux of the \emph{unextincted} template found in the $\chi^2$ analysis
averaged over a 200 $\AA$ band centered on 1500 $\AA$. The reshift and
the normalization factor A (equation~\ref{eq:a}) set the absolute 1500
$\AA$ flux. This procedure provides the extinction corrected 1500 $\AA$
flux for the SFR calculation.  It is obviously a simplification to
assume a single extinction for a whole galaxy but it is clearly better
than assuming a single average extinction for all galaxies or no
extinction at all.  A much more extensive discussion of the extinction
is presented in TWS and it is not repeated here.

Rather than using the isophotal or total flux values returned by SE,
the flux in each band is computed as the sum of the flux from each
pixel that the Szalay procedure and SE determine is part of an object.
The SFR computed with the A value determined from the $0.6 \arcsec$
aperture is then multiplied by the ratio of the total flux to the
aperture flux to give the values reported in Table~\ref{tab1}.  The sum
of the SFRs in Table~\ref{tab1} gives the observed SFR in the
appropriate redshift bins. At high redshift the observed SFR is
significantly below the actual SFR due to surface brightness dimming.
The correction for this effect, using the extinction corrected star
formation rate per pixel is described in \S~\ref{sec-sbd}.

\section{Correction for Surface Brightness Dimming} \label{sec-sbd}

It is well known that surface brightness dimming at high redshifts
limits the number as well as the extent of galaxies that are detected,
creating errors in the measured SFR.  In TWS we used the star formation
intensity distribution function \citep{lanz99} to correct for star
formation missed due to surface brightness dimming.  \citet{lanz02} has
also used the distribution to correct for missed star formation but did
not use a distribution derived from extinction corrected UV fluxes.

\subsection{Star Formation Intensity Distribution Function}

The star formation intensity x is defined as the SFR in solar masses
per year per \emph{proper} square kiloparsec. The intensity is
calculated for each pixel that is part of a galaxy.  Within a given
redshift interval the distribution function, h(x), is defined as the
sum of all the proper areas in a interval of specific intensity,
divided by the interval and by the comoving volume in cubic megaparsecs
defined by the field and redshift interval \citep{lanz99}.  Defined in
this manner the values of h(x) determine the star formation rate per
cubic comoving megaparsec through equation(\ref{eq:hxsfr}).

\begin{equation} sfr = \int_{0}^{\infty} xh(x)dx \label{eq:hxsfr}
\end{equation}

\begin{figure}

\plotone{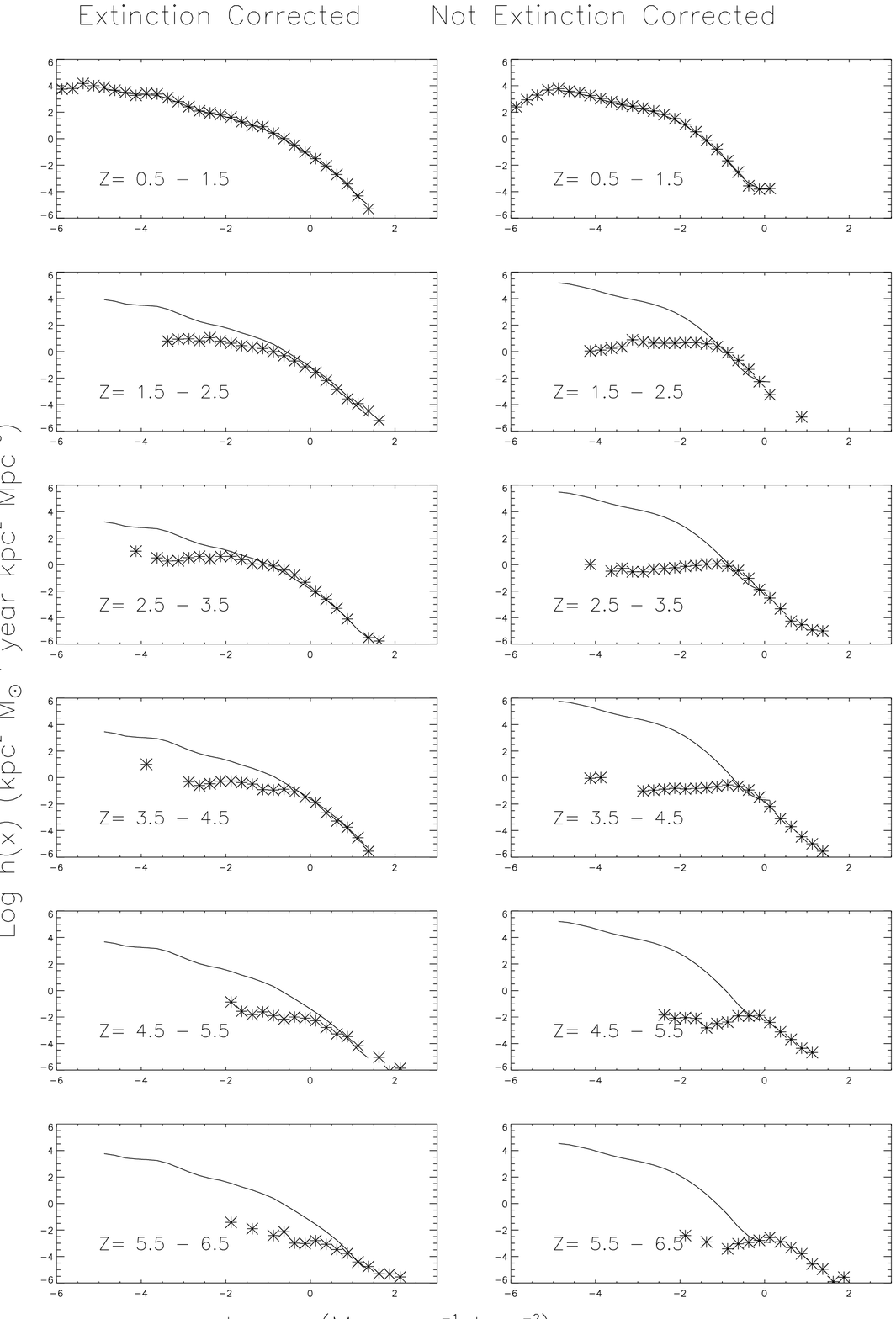}

\caption{Plots of the SFR intensity distribution for the Northern HDF.
The abscissa is the log of the SFR per unit area while the ordinate
gives the distribution of this quantity per comoving volume in the
redshift bin. The left hand panel is for SFRs corrected for dust
extinction, while in the right hand panel no extinction correction has
been applied. The solid lines are the smoothed functions in the
redshift 0.5 - 1.5 bins matched to the high intensity regions of the
distribution at higher redshifts.  See text for further discussion.}

\label{fig-hx}
\end{figure}

The distribution functions for both extinction corrected and uncorrected
star formation intensities are shown in Figure~\ref{fig-hx}.

\clearpage

The top panels of Figure~\ref{fig-hx} show the 3 point smoothed fit to
both the extinction corrected and uncorrected distribution function at
a redshift of 1.  At higher redshifts the smoothed curve is matched to
the high intensity portion of the distribution by simply moving the
smoothed distribution up or down in the log-log plot.  It is clear that
the smoothed curve for the uncorrected distribution does not match well
at higher redshifts.  This is discussed further in \S~\ref{ssec-val}
but for now we will concentrate only on the extinction corrected
distribution.

The deviation of the observed data from the smoothed distribution in
the lower panels shows the effect of surface brightness dimming.  As in
TWS we correct for surface brightness dimming by integrating
equation(\ref{eq:hxsfr}) over the observed data where the smoothed
function matches the data and then over the smoothed function (adjusted
in height) when the observations fall below the curve.  Note that the
integral reaches 59\% of its total value at $\log{x} = -0.25$ and 94\%
of its total value at $\log{x} = -1.25$. \S~\ref{ssec-val} discusses the
validity of the correction.

Table~\ref{tab-sfr} shows the SFR at each stage of correction.  The
second column gives the rate with no correction for either extinction
or surface brightness dimming.  The third column shows the SFR after
correction for extinction only and the fourth column shows the SFR
after correction for both extinction and surface brightness dimming.
At low redshifts extinction is the dominant correction while at high
redshifts extinction and surface brightness dimming are comparable.
All of the calculations without extinction correction are made from the
output of the $\chi^2$ photometric redshift program with the extinction
held to zero.  In some cases this alters the photometric redshift since
the degree of freedom to redden a galaxy by extinction is removed.
This is the cause of the odd case at redshift 4 where the extinction
corrected SFR is less than the uncorrected SFR.

\subsection{Selection Function} \label{ssec-sfun}

\citet{lanz02} introduced the concept of a selection function in
conjunction with the star formation intensity distribution function
h(x).  The selection function is the angular area over which a
parameter at a given depth can be detected.  The maximum angular area
is the total area of the observation. At some depth of a parameter such
as flux, the area goes to zero when it becomes undetectable.  The
parameter used in \citep{lanz02} and here is the star formation
intensity x.  The main reason for introducing this concept is the
variance in sensitivity of the NICMOS Camera 3 in different regions of
the detector.  The sensitivity varies by a factor of 3 from maximum to
minimum.

The star formation intensity is a function of redshift, SED,
extinction, and flux, therefore, each pixel included in h(x) at a
specific x may have a different selection function depending on the
values of the parameters.  Given the criterion of 3.5 sigma detection
in a flux band for 3 contiguous pixels the selection function
determines for each source pixel the percentage of the field area where
the source could be detected.  The star formation intensity
distribution function program then corrects the area of each pixel by
the inverse of the selection function.  The maximum allowed correction
is a factor of 10 to prevent marginal detections from dominating the
distribution function. In practice only a small percentage of pixels
have selection functions less than 100\%.

\subsection{Validity of the Distribution and Comparison With Other Work} 
\label{ssec-val}

Initially in TWS the star formation intensity distribution was treated
strictly empirically with no physical motivation.  The identical match
of the smoothed distribution derived here from largely independent data
is further empirical confirmation of the distribution as is the
excellent match at high SFR intensities at higher redshifts.
\citet{thm02}, however, provides a physical motivation by showing that
the distribution is a natural consequence of the Schmidt law, the
majority of star formation in  occurring in exponential disks, and a
Schechter distribution of galaxy masses. It also showed that effects of
a smaller characteristic mass and a smaller average radius of the
exponential disk at higher redshift have opposite effects on the shape
of the distribution leading to a generally invariant distribution.
Although we lack detailed knowledge of the changes of characteristic
mass and exponential disk radius with redshift, the assumption of an
invariant distribution appears to be justified. Even if the surface
brightness dimming corrections in Table~\ref{tab-sfr} varied by 30\%
the error would be small compared to the errors from other effects such
as photometric error and large scale structure.

\citet{lanz02} also utilized the star formation intensity
distribution function to correct for surface brightness dimming but
found a SFR that steadily increases from the present day to high values
at a redshift of 8, in contrast to the essentially steady star formation
rate between a redshift of 1 and 6 as presented in \S~\ref{sec-sfh}.
Only the SFR in the common ground of redshifts will be discussed in the
following comments.  The comparison will be with the rates found from
scaling the h(x) distribution vertically in \citet{lanz02} since that
is the technique used in this paper.

The critical difference between the technique applied here
versus \citet{lanz02} is that this work uses the extinction corrected
distribution where as \citet{lanz02} use the distribution that is
uncorrected for extinction.  \citet{lanz02} refer to this rate as the
unobscured rate.  At a redshift of 1 they find a SFR of
$0.03 M_{\sun}$ per Mpc$^{-3}$ whereas the value in this work is $0.3M
_{\sun}$ per Mpc$^{-3}$.  This is simply understood as the
difference between correcting for extinction and not correcting for
extinction.  At a redshift of 4 \citet{lanz02} find a rate of $0.25
M_{\sun}$ per Mpc$^{-3}$ comparable to our value of 0.1.  At a redshift
of 6 they find a value of 0.45 compared to our value of 0.2.

The difference between our relatively steady values and their
increasing values is due to two effects.  As can be seen from
Figure~\ref{fig-hx} and \citet{lanz02} the bright end of the
distribution has a steeper slope in the uncorrected for extinction case
than the corrected one.  This is due to extinction removing
intrinsically bright pixels from the distribution.  Second at higher
redshifts the average extinction of the sample becomes smaller due to
surface brightness dimming removing the highly extincted galaxies from
the sample.  Matching the steep slope of the uncorrected distribution
function to the continuously lower extinction sample at higher
redshifts results in an over correction and an apparent steadily
increasing star formation rate.  Figure~\ref{fig-hx} shows this effect.
The dimming correction in the right hand column is significantly higher
than in the left hand extinction corrected column.

\section{Star Formation History} \label{sec-sfh}

Initial work on the star formation history of the universe by
\citep{mad96} showed a sharp rise in SFR from the present day to a peak
at z between 1 and 2 then a decline to lower rates at higher
redshifts.  Later work by \citet{mad98}, which applied a small
correction for extinction, showed a shallower decline in SFR at high
redshifts. Subsequent work by \citet{sti99}, \citet{hop00} and TWS
gave SFRs that were roughly constant at redshifts higher than 1. This is
consistent with the results presented here.  The SFR values found in
this paper are also consistent with the SFRs found in the
sub-millimeter observations of \citet{bar00} at redshifts of 1-3 and
3-6. The concordance of results between the various studies is
discused in \S~\ref{s-con}.

\begin{figure}

\plotone{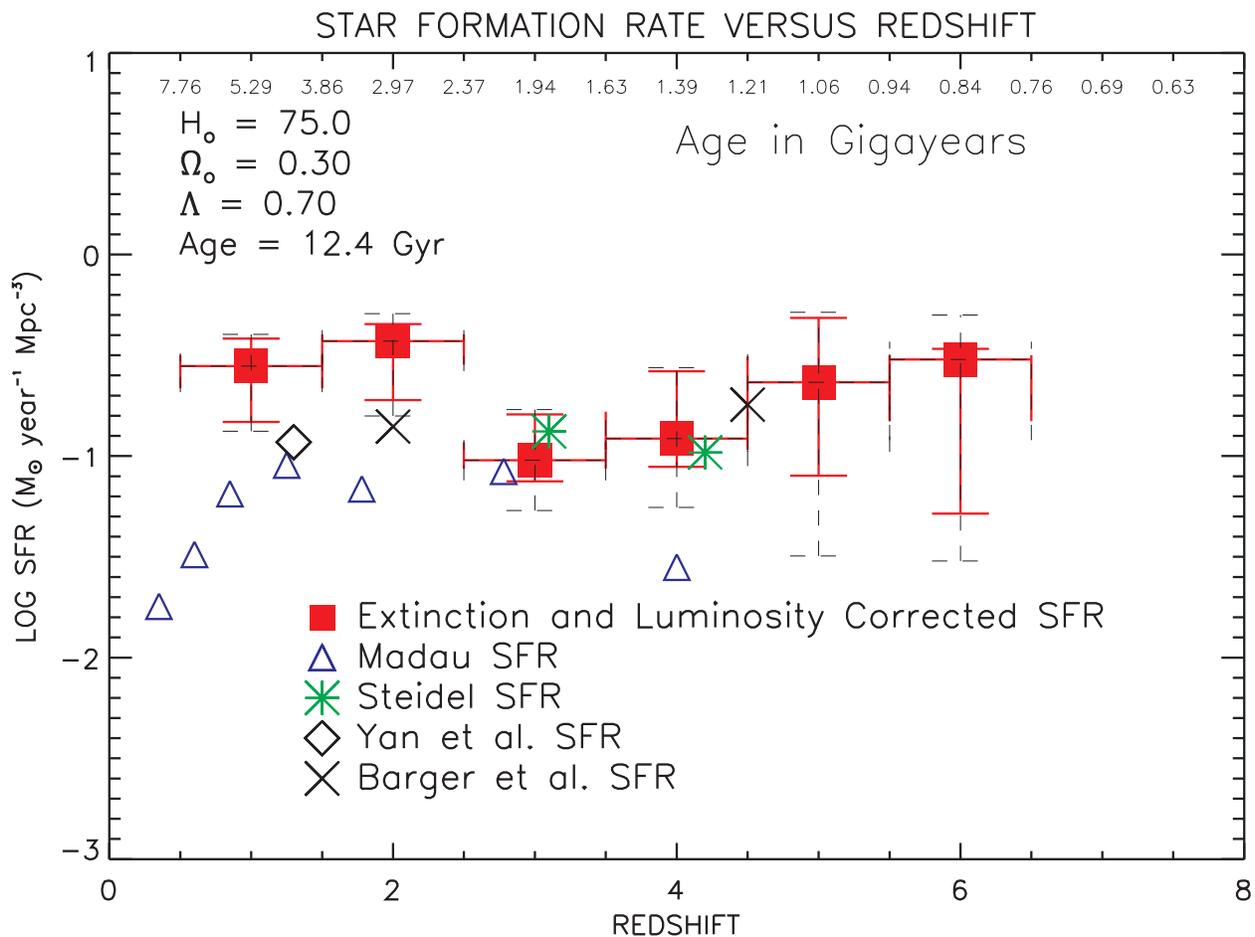}

\caption{Plots of the extinction and surface brightness dimming
corrected SFR as a function of redshift.  The solid error bars indicate
the photometric errors and the dashed error bars indicate the
uncertainty in the global SFR.  The SFR from \citet{yan99} indicated by
the diamond is for star formation in the range between a redshift of
0.8 to 2.0. The SFRs from \citet{bar00} are indicated by the crosses
and are for redshifts 1-3 and 3-6 with a $\Lambda$ = 2/3 universe.}

\label{fig-sfr}
\end{figure}

Figure~\ref{fig-sfr} shows the SFR versus redshift corrected for the
surface brightness dimming and the selection function as described in
\S~\ref{sec-sbd}.  The redshift binning unit is 1 centered on integer
redshifts.  The lowest redshift included in the study is 0.5 and the
highest is 6.5.  The NHDF statistics are not adequate at redshifts
below 0.5 to determine an accurate SFR.  In that redshift range the
number of sources is low and the percentage error in redshift is high,
leading to high SFR errors.  The accuracy of large area ground based
surveys is much greater in this redshift range.  Although
Figure~\ref{fig-hmag} shows galaxies at redshifts above 6.5 we do not
use them in this analysis.

  \subsection{Star Formation Time History} \label{ssec-sft}

\begin{figure}

\plotone{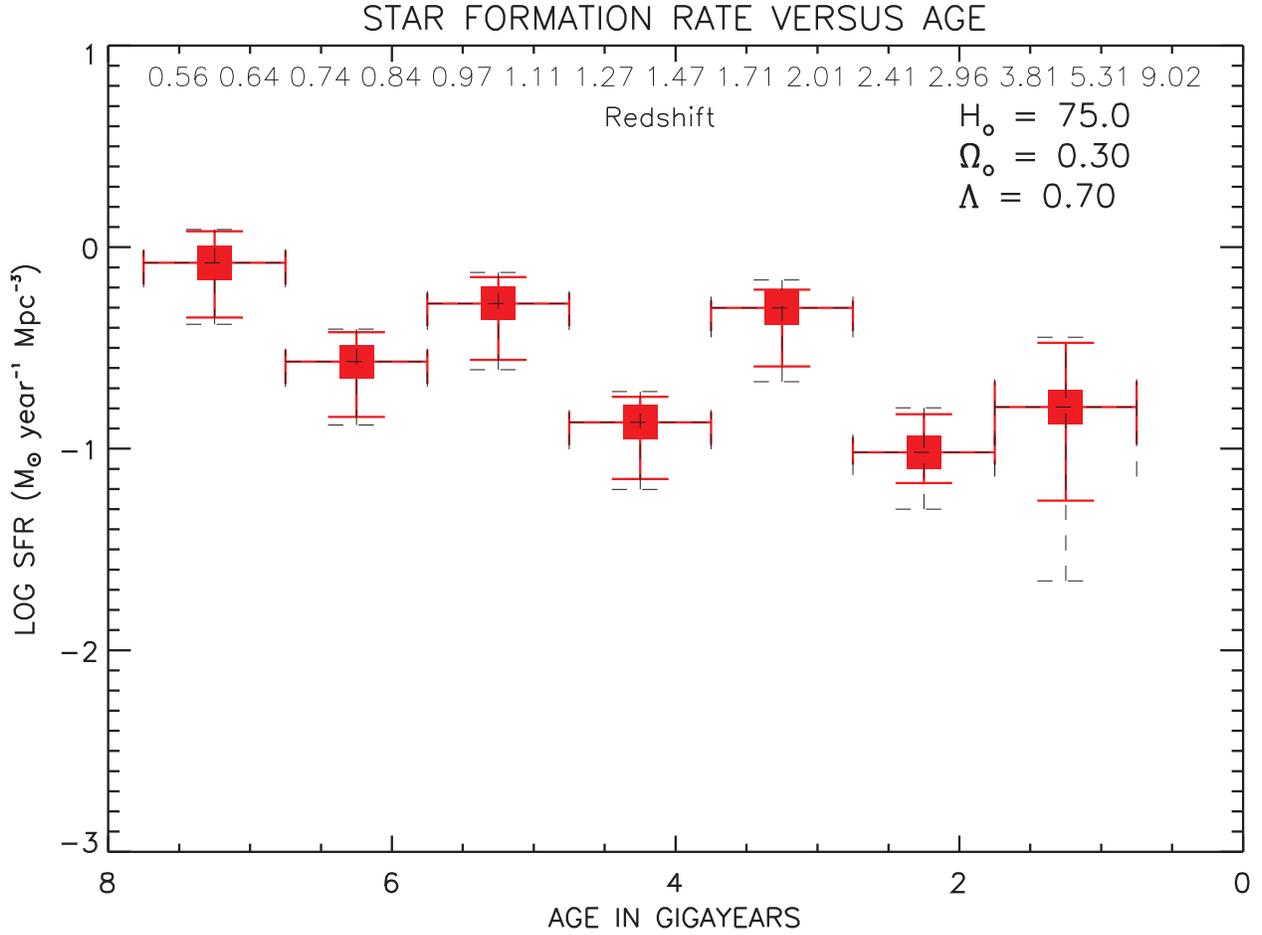}

\caption{Plots of the extinction and surface brightness dimming
corrected SFR as a function of age of the universe for the cosmology
indicated in the figure.  The solid error bars indicate the photometric
errors and the dashed error bars indicate the uncertainty in the global
SFR.}

\label{fig-sfrt}
\end{figure}

The usual form of the `Madau Diagram' plots SFR versus redshift since
redshift is the measured quantity and is not subject to the choice of
cosmology.  This form of the diagram can be misleading in terms of when
the majority of star formation occurs since there is very little time
at high redshifts.  A more instructive plot is the star formation rate
versus age of the universe.  In the era of `precision cosmology' one
can dare to produce such a plot with reasonable assumptions on the
cosmology.  For ages in the redshift range of 0.5 to 6.5
Figure~\ref{fig-sfrt} indicates the measured SFR at ages of the
universe between 1.25 and 7.25 gigayears.  The non integer age units
are dictated by the ends of the redshift range.

Within the errors the SFR is roughly constant from an age of the
universe between 1 and 7 gigayears at a rate significantly above the
present day rate.  The trend of the data indicates an increase in the
formation rate from 1 gigayear to 7 gigayears but the errors are too
large to validate this impression.  Note that the binning of the
objects is quite different between the SFR versus redshift and versus
age of the universe plots.

\subsection{Error Analysis}

The error analysis for this data set was carried out in the same manner
as the analysis performed in TWS where there is an extensive discussion
of the analysis procedures. Table~\ref{tab-ers} gives the error sources
from number statistics, photometric error and large scale structure.
The greatly increased number of sources in this data set has reduced
number statistics to a negligible error compared to other error
sources. 

\subsubsection{Photometric Error Propagation}

Photometric errors can propagate into errors in redshift, extinction,
and SED.  These in turn create errors in the SFR.  As in TWS, the
photometric error analysis computes the SFR in each redshift bin for
100 runs of the data with the flux values of each object altered
randomly by the gaussian distribution of the flux errors calculated for
each object.  The flux error is the total flux error which includes
both the random and systematic error shown in the denominator of
equation~\ref{eq:chisq}. Due to the variation of sensitivity across the
NICMOS detector, the errors in NICMOS flux are calculated individually
for each galaxy.  The distribution of SFRs between the runs determines
the SFR error level from photometric errors. The solid line photometric
error bars in Figure~\ref{fig-sfr} are the 16\% and 84\% points of the
SFR distribution in quadrature with the 17\% redshift error discussed
in \S~\ref{ss-prac}.  This procedure integrates the photometric error
induced redshift, extinction and SED errors into a measured SFR error.

\begin{figure}

\plotone{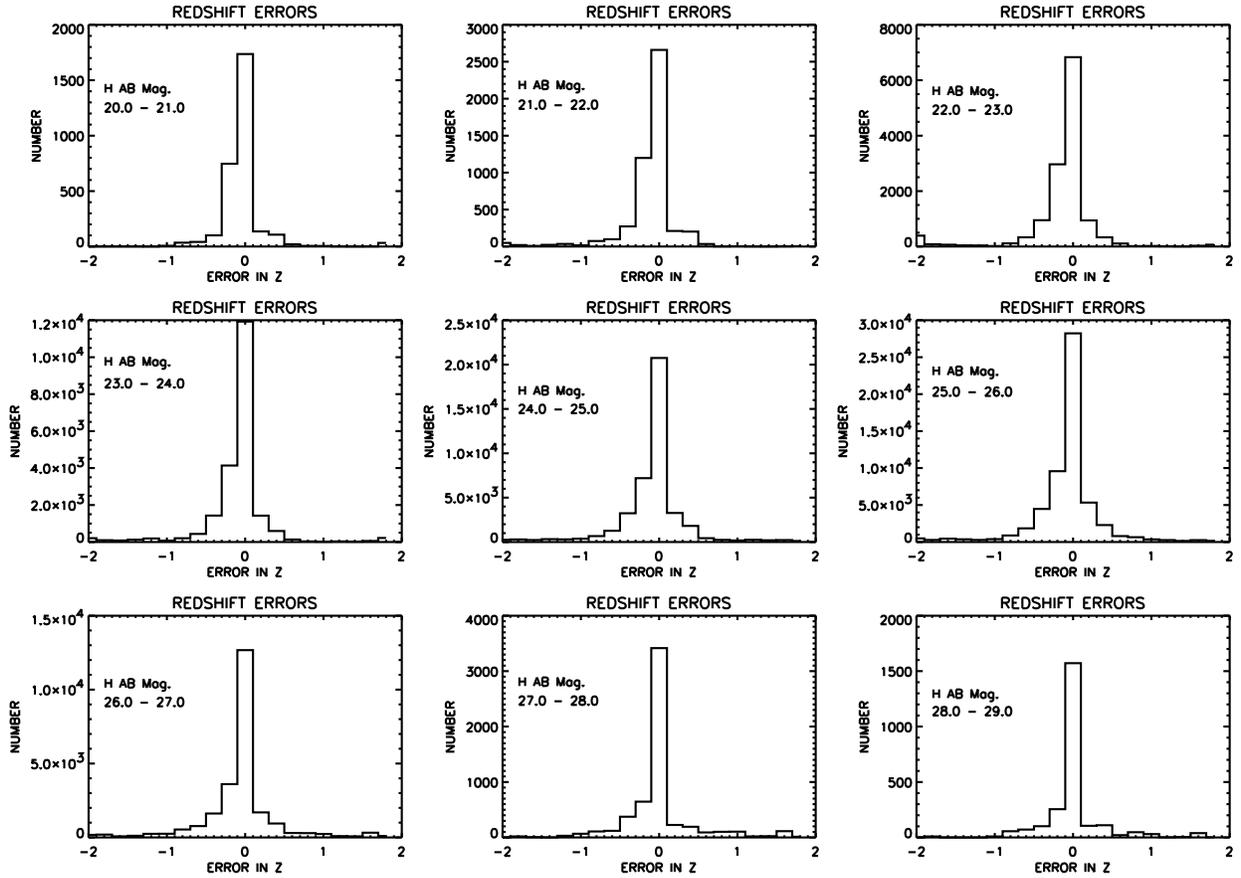}

\caption{The redshift error histograms for all sources in
Table~\ref{tab1}. Negative values indicate that the unperturbed
redshift is higher than the perturbed redshift.}

\label{fig-zer}
\end{figure}

\begin{figure}

\plotone{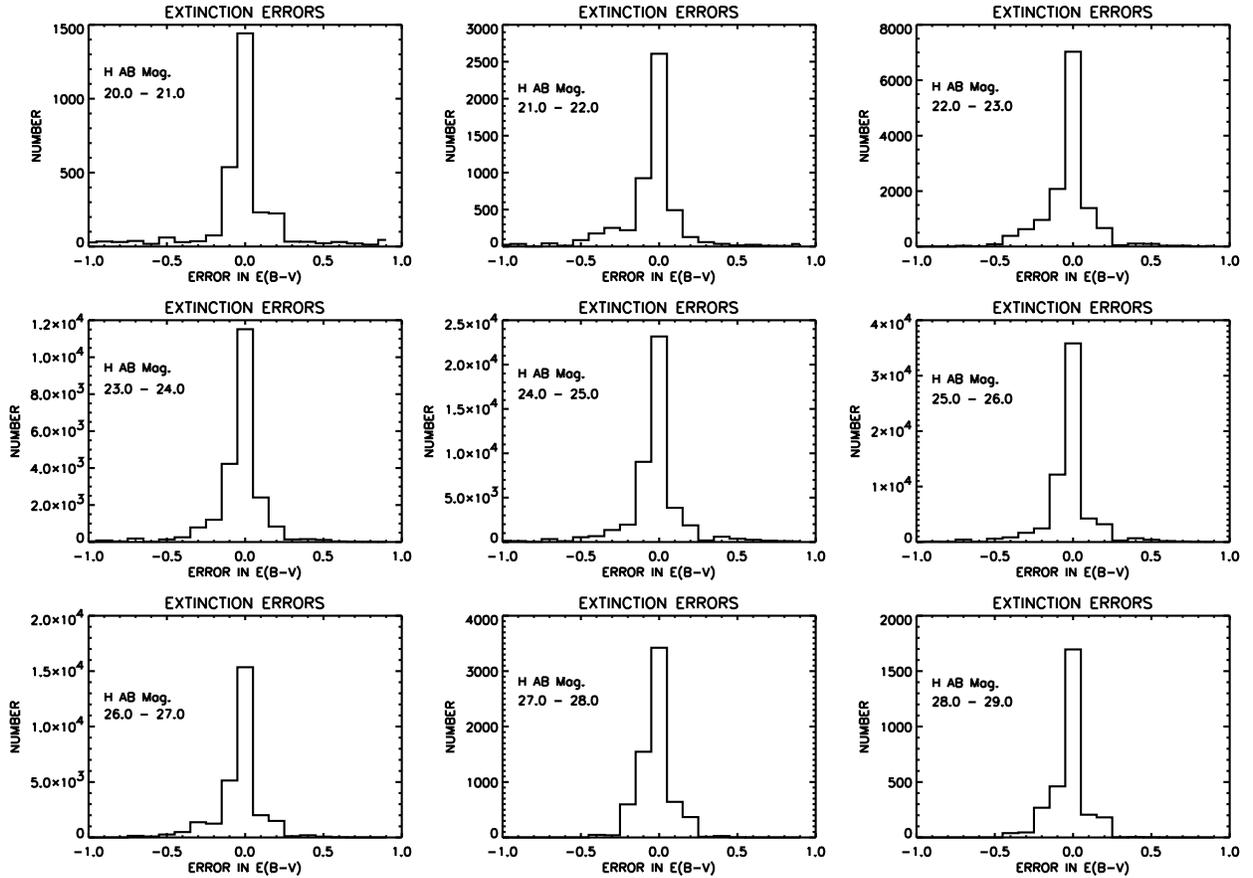}

\caption{The E(B-V) error histograms for all sources in Table~\ref{tab1}.
Negative values indicate that the unperturbed extinction is higher than
the perturbed extinction.}

\label{fig-eer}
\end{figure}

\begin{figure}

\plotone{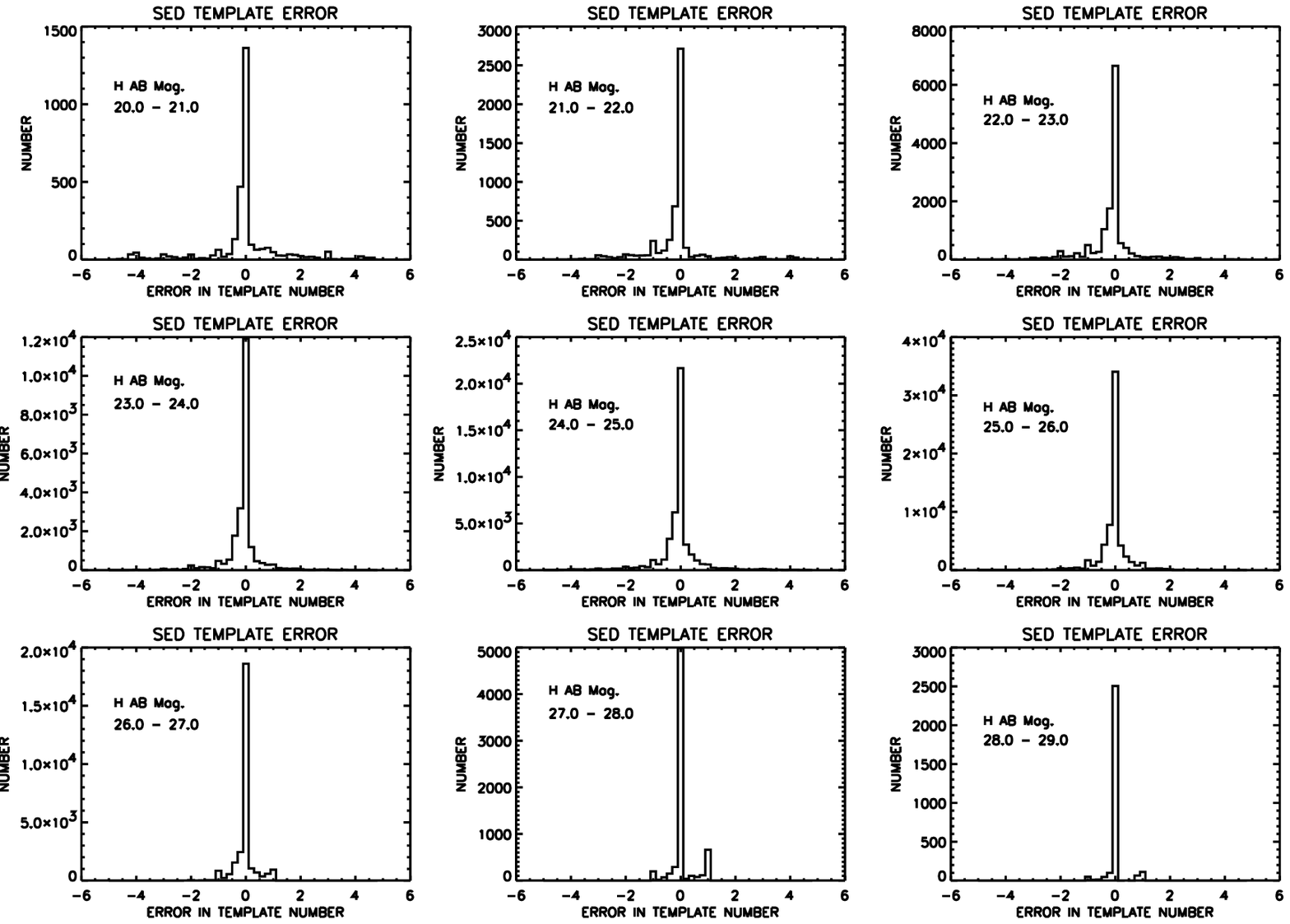}

\caption{The template number error histograms for all sources in
Table~\ref{tab1}. Negative values indicate that the unperturbed
template SED is later than the perturbed SED.  Template numbers run
from 1, the earliest, to 6, the latest in increments of 0.1}

\label{fig-ter}
\end{figure}

The procedure preserves the redshift, extinction, and SED calculated in
each run, therefore, we can plot the individual errors in each
parameter as a function of the F160W AB magnitude.  Figures
\ref{fig-zer}, \ref{fig-eer} and \ref{fig-ter} show the histograms of
redshift, extinction and template SED errors in F160W AB magnitude bins
ranging from 20 to 29 for all of the sources listed in
Table~\ref{tab1}. In each case the unperturbed value is subtracted from
the perturbed value.  Note that since the histograms include the more
error prone objects in the 0.0-0.5 and 6.6-8.0 redshift ranges they are
a conservative measure of the error.

The error histograms do not show any significant trend with magnitude.
The basic reason is that the systematic part of the error scales with
flux so that the perturbations for the strong fluxes are larger than
for the weaker fluxes, accurately reflecting the observations.  As
mentioned in \S~\ref{s-pres}, the 10\% systematic error is probably an
overestimation so errors at brighter magnitudes may be overestimations.
If this is the case the error bars in Figure~\ref{fig-sfr} may be
exaggerated.

Both the redshift and extinction errors show a trend of being asymmetric
around zero error with an over abundance of negative errors showing
larger redshifts and extinctions for the unperturbed values than for
the perturbed values.  Both of these effects indicate larger SFRs for
the unperturbed versus the perturbed values.  This trend is reflected
in the overall SFR errors shown in Figure~\ref{fig-sfr} where the error
bars extend further on the low side than the high side for most of the
redshift bins.

\subsubsection{Large Scale Structure Error}

As in TWS the two-point correlation function $ \xi(r_o,\gamma) = (r_o/r)^{\gamma}$ as
given by \citet{peb80} is used to calculate the large scale structure error.  The 
fractional error is taken as $\sigma_N/N$ where

\begin{equation}  \sigma_N/N  =   \sqrt{ 1 + N  \times I_2} / \sqrt N 
\label{eq:stat1} \end{equation}

\noindent and  

\begin{equation}  I_2 = { \int \int  \xi(r_o,\gamma) dV_1 dV_2}/V^2  
\label{eq:stat2}
\end{equation}

For each redshift bin it is a very good approximation to
consider the volume as a long thin tube of square cross section with
sides of (comoving) dimension D and (comoving) length L.  Then the
expression for $I_2$ has the form

\begin{equation}  I_2 \simeq C(\gamma) \times  (r_o/D)^\gamma \times (D/L) 
\label{eq:stat3}
\end{equation}

The quantity $L/D$ is simply the number of cubes of dimension $D\times
D\times D$ that can be placed end to end in the tube of length L. The
dimensionless coefficient $C(\gamma)$ is a double integral in which
$dV_1$ is taken over the unit cube and $dV_2$ covers that same cube
plus a large number of cubes on either side of this unit cube. We have
evaluated this coefficient numerically which is vastly simplified by
the large number of symmetries in this geometry. From the work of
\citet{ade98} we adopt $\gamma$ = 1.8 and $r_o = 5 h^{-1}$ Mpc and find
$C(\gamma = 1.8)$ = 8.22. We use the value of D and L for our adopted
cosmology appropriate to the center of each bin.

The calculation was made for a square field equal to one WFPC2 chip and
then the errors for the three fields were added in quadrature.  The
larger field and the increased photometric error means that large scale
structure is no longer the dominant but still an important error in
extrapolating the SFR history in the NHDF to the SFR history of the
universe. The dashed error bars in Figures~\ref{fig-sfr}
and~\ref{fig-sfrt} are the quadrature sum of the photometric and large
scale structure errors.

\section{Magnitude Functions and Comparisons with Predictions} 
\label{sec-lf}

\begin{figure}

\plotone{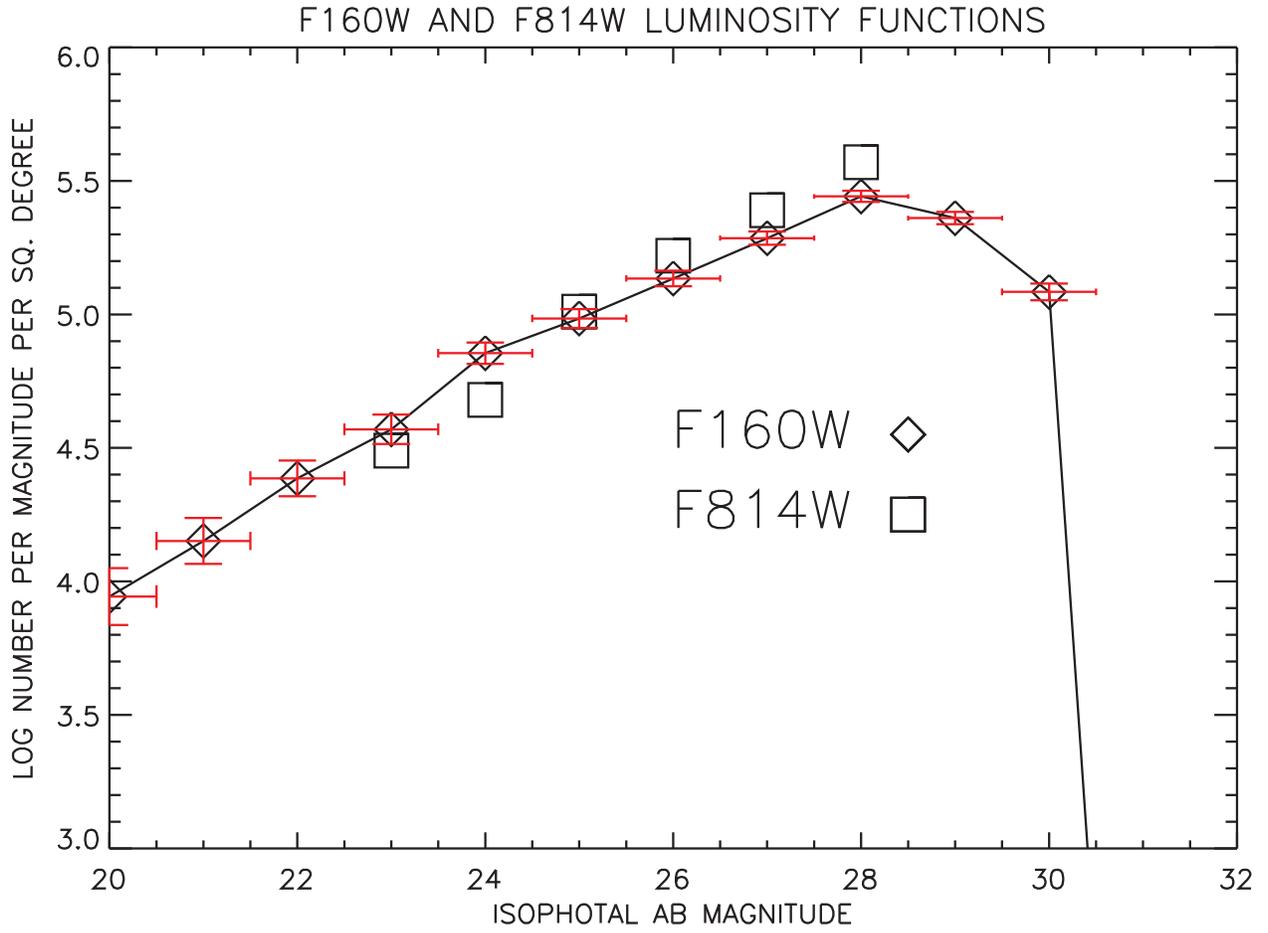}

\caption{The NICMOS F160W (1.6 micron) magnitude function with the 
WFPC2 0.814 micron magnitude function superimposed.}

\label{fig-hlum}
\end{figure}

Ground based K magnitude functions have provided constraints on galaxy
formation models, eg. \citet{kaf98}.  The much deeper F160W magnitude
function determined here provides an even more powerful constraint on
theoretical models.  A great advantage of this function is that is
completely independent of models or cosmology.  Figure~\ref{fig-hlum}
shows the observed F160W magnitude function for the NHDF.  The roll off
at magnitudes greater than 28 is due to surface brightness dimming and
is not a true feature of the magnitude function.  In the range of
validity, the magnitude functions from the WFPC2 F814W band
\citep{wil96} and the NICMOS F160W band are remarkably similar.

\cite{kaf98} proposed two tests to discriminate between hierarchical
galaxy formation and pure luminosity evolution (PLE) of galaxies. Both
models were constrained to fit the present day K band luminosity
function.  Since the predictions are for the K band, K band magnitudes
were calculated from the observed 1.6 micron flux using the SEDs found
by the analysis. Any error in this transformation is very small
compared to the differences in the predictions between the hierarchical
and PLE models. The first test compares the number density of galaxies
versus K magnitude at redshifts of 0.6, 1.0, and
2.0. The PLE model predicts many more bright galaxies at each redshift
than do the hierarchical models.  Since the predictions only go as
faint as a K magnitude of 21.5 the number statistics in the NHDF are
small.  The PLE model predicts a factor of 10 more bright galaxies than
are observed in the NHDF at redshifts of 1 and 2.  The hierarchical
predictions are within the error bars of the observations.

\begin{figure} 

\plotone{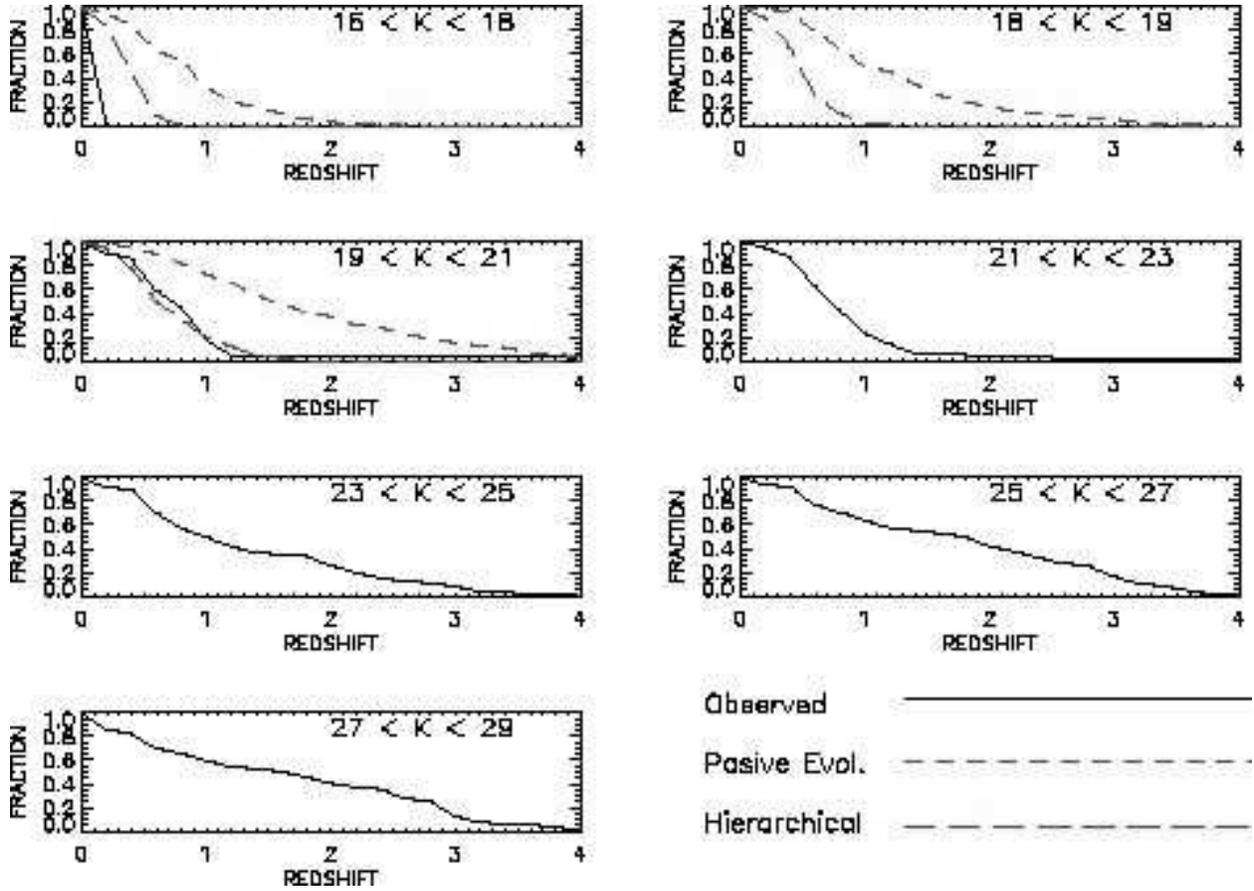}

\caption{Plots of the relative fraction of galaxies with redshifts
greater than z in 7 different magnitude bins. Predictions only exist
for the first three bins.  The first two bins have only 1 and 0
observed galaxies respectively.} 

\label{fig-frt}

\end{figure}

A more decisive test from \cite{kaf98} is their prediction of the
relative fraction of galaxies versus redshift in K magnitude bins.
Figure~\ref{fig-frt} shows the predictions and the observations.  The
first two magnitude bins have 1 and 0 galaxies in the NHDF but the
third K 19-21 bin has a significant number of galaxies.  It clearly
favors the hierarchical model with the data and prediction being
virtually indistinguishable.  Although the predictions end at a K
magnitude of 21, the data for fainter magnitudes follows an
extrapolation of the hierarchical prediction at those magnitudes.   The
net results of the two tests are consistent with the hierarchical
prediction and inconsistent with the PLE prediction.

\begin{figure}
 
\plotone{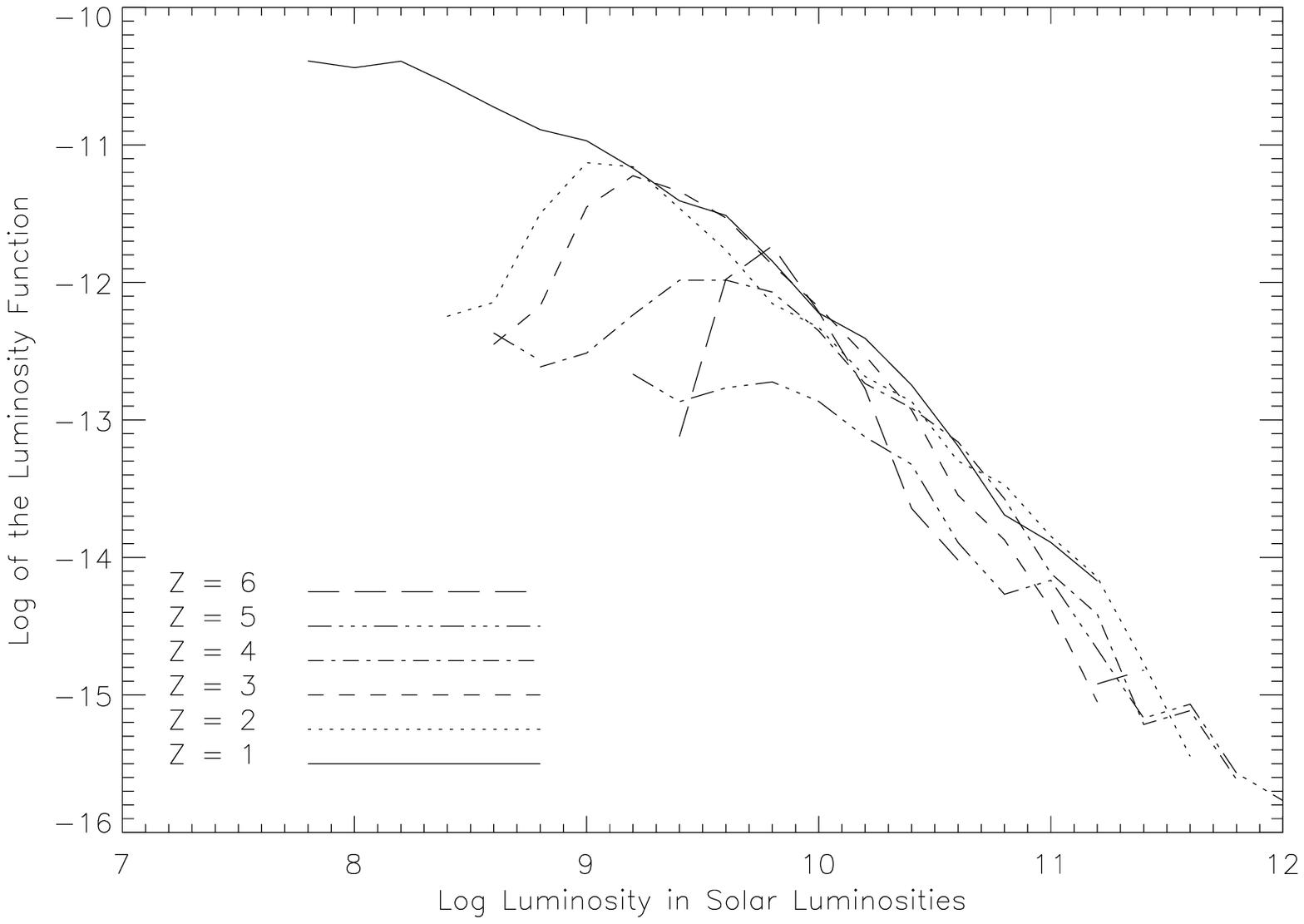}

\caption{The evolution of the 1.6 micron luminosity function with 
redshift.} 

\label{fig-lume}

\end{figure}

Our analysis determines the redshift, extinction, and intrinsic SED
therefore a combination of the flux, redshift, and SED determines the
total luminosity of the galaxy to the accuracy of the individual
parameters.  The great depth of NHDF observations provides further
constraints on galaxy formation models by observing the evolution of
the luminosity function with time.  Figure~\ref{fig-lume} shows the
luminosity function at 6 different epochs, z = 1-6.  The roll off of
each curve at low luminosities is simply due to galaxies falling below
the detection limit.  To observational accuracy, the shapes of the
luminosity functions at each epoch appear remarkably similar.  The low
absolute values  for the luminosity function at a redshift of 5 may be
due to a void in the NHDF at that redshift.

It should be noted that even though the luminosity function appears
invariant with redshift to z = 6, this does not mean that the mass
function is also invariant.  Very young galaxies at a redshift of 6
require much less mass to achieve the same luminosity as old, present
day galaxies.  In fact, the invariant luminosity function appears to
favor the hierarchical model of galaxy assembly.

\section{Infrared and Sub-mm Constraints} \label{sec-ir}

The observed infrared and sub-mm backgrounds place constraints on the
extinction values found in our $\chi^2$ analysis.  If the extincted
luminosity is significantly different from the observed infrared
background levels then the extinction values may be considered
suspect.  A calculated background flux that is much higher than the
observed background would indicate an over correction for extinction
and a higher SFR than allowed by the infrared and sub-mm constraints.
On the other hand, if the extincted luminosity is much lower than the
infrared background levels it would be a sign that our analysis has
missed a large part of the highly obscured star formation activity.  In
addition the calculated sub-mm luminosity of individual sources should
not exceed the detections and limits set in the NHDF by \citet{hug98}.
Note that although comparison with the observed background and emission
from individual objects is a constraint on the extinction values,
agreement is not a validation of the values.  The difference between
E(B-V) values of 1 and 2 is a large extinction error but both predict
essentially the same amount of re-emitted flux since they both remove
almost all of the UV and optical luminosity.

Both \citet{cal00} and \citet{ade00} have explored the errors in
transforming between extinctions derived from the optical flux and the
observed infrared emission. \citet{cal00} estimates that the error is
a factor of 2 for individual sources which reduces to 20\%
for a group of about 50 galaxies. \citet{ade00} cite errors of 0.2 -
0.3 dex (1.6 - 2.0) and ascribe the same range in SFR error.
The error bars in Figures~\ref{fig-sfr} and~\ref{fig-sfrt} are well
within these limits, particularly since the points represent
hundreds of galaxies rather than individual galaxies.

\subsection{Predicted and Observed Backgrounds}  \label{ssec-bac}

The fraction of luminosity removed by extinction is easily calculated
for each SED from the  Calzetti obscuration law.  This luminosity is
assigned to a dust emission SED.  Dust SEDs vary significantly
\citep{rig99} but there is no easy way to predict which SED is
appropriate from just the optical and near-IR observations.  For
consistency a single dust SED is used from Figure 1 of \citet{aus99}
for the PAH features joined with the Arp 220 ULIRG spectrum given in
Figure 4 of \citet{rr93}. The SCUBA 850 \micron\ flux as well as the
ISO 6 and 15 \micron\ fluxes listed in Table~\ref{tab1} are calculated
from the SED, E(B-V), redshift and luminosity.

Integration of all of the predicted 850 \micron\ fluxes gives a 850
\micron\ background of $3.9 \times 10^{-10}$ watts m$^{-2}$
steradian$^{-1}$.  The 850 \micron\ flux is consistent with the value
of $5 \pm 2 \times 10^{-10}$ from \citet{bla99} and $4 \times 10^{-10}$
for COBE \citep{fix98}.  The agreement with the probably more accurate
COBE flux is remarkable but it should be interpreted only that the
predicted sub-mm flux does not indicate an error in the derived
parameters

Integration of the observed 1.6 \micron\ fluxes gives  a 1.6
\micron\ discrete source background of $7.0 \times 10^{-9}$ watts
m$^{-2}$ steradian$^{-1}$.  The previous 1.6 \micron\ background in the
Deep NICMOS NHDF from TWS was $6 \times 10^{-9}$.  The F160W background
found here is significantly less than the J ($54 \times 10^{-9}$) and K
($27.8 \times 10^{-9}$) backgrounds found by \citet{cam01}. It is also
less than but closer to the K band background of $12.4 \times 10^{-9}$
found by \citet{wrt00}.  Our background subtraction technique will
remove any component of the true background not due to detected
discrete sources.

\subsection{Individual Sources}

Although it is not the primary purpose of the infrared flux
calculation, the calculated fluxes of individual sources can be
compared with observations and observational limits.

\subsubsection{ULIRGs} \label{sss-ulirg}

We define an Ultra Luminous Infrared Galaxy (ULIRG) as a galaxy with
greater than $10^{12}$ L$_{\sun}$ total luminosity with half or greater
of the luminosity emitted in the mid and far infrared.  One of the
surprising aspects of TWS was the observation of two ULIRGs  in the
small 50$\arcsec$x50$\arcsec$ field of the Deep NICMOS observations.
They are sources 439 and 800 in the present catalog.  Source 800
retains it ULIRG status in this analysis.  Source 800 may also contain
an AGN component.  Although not listed in their detection catalog, the
Chandra X-Ray map of \citep{bra01b} clearly shows a source at the
position of source 800. Source 439 (166 in TWS), however, falls to LIRG
status with a luminosity of $1.95 \times 10^{11}$ L$_{\sun}$.  The
primary reason is that the best fit in this analysis is a template 5.7
and E(B-V) = 0.5 fit as compared to a template 5.9 and E(B-V) = 1.0 fit
in TWS.  Although the differences in predicted fluxes is small, the
luminosity difference between the two fits is large.  Given the higher
signal to noise, the parameters found in TWS are more likely correct.

There are two very high redshift galaxies, 22 and 396, with ULIRG level 
luminosities but their redshifts of 8.0 and 7.44 put them beyond the galaxies 
considered in this paper.  The validity of these source will be left to 
another study. An additional two candidates included in this analysis
are sources 475 and 1364 at redshifts of 6.24 and 4.88 respectively.
These do not correspond to any WFPC2 NHDF source and are both outside
of the field of the Deep NICMOS image.  Source 1364 has a small amount
of F814W flux while 475 has no detectable flux in any of the optical
bands.  Both galaxies have significant F110W and F160W fluxes with very
red F160W-F110W colors. The $\chi^2$ plots show two very broad minima
at redshifts between 1 and 2 and at the higher redshifts where the fit
is the best. Both of the high redshift minima are extremely broad,
covering a redshift range of approximately 2 around the selected
redshift.  Given the nature of the $\chi^2$ distribution, the ULIRG
status of these sources is extremely uncertain.

With no reliable detection of additional ULIRGs in this study we can
throw no light on the question of whether ULIRGs were much more
prevalent in the past, reaching a peak near a redshift of two.  If the
two possible ULIRGs above were at a redshift of 2 rather than their
high redshifts, their luminosities would be considerably sub-ULIRG.
The existence of any ULIRGs in the area of the NHDF is still highly
improbable, however, given the local space density of ULIRGs.
Resolution of this question will require deep imaging in additional
fields.

\subsubsection{SCUBA Sources} \label{sss-scuba}

Probably the most enigmatic sources in the NHDF are the sub-mm sources
detected by SCUBA \citep{hug98}. Of the 5 detected HDF sources, source
2 actually lies just outside of the NHDF field and source 3 is in the
PC chip which was not analyzed in this work. The remaining sources
1, 4, and 5 are discussed below.

VLA observations by \citet{dwn99} indicate that the galaxy
corresponding to HDF850.1 is either WFPC2 3-586 or 3-593 if it is
associated with an optically visible galaxy.  It could also be an
optically obscured galaxy lying between the two galaxies.  Examination
of the F160W image does not reveal an optically undetected galaxy lying
between the sources, however, the images of the two galaxies slightly
overlap at a level below the threshold set for positive detection in
\S~\ref{sec-dr}. WFPC2 3-586 corresponds to source 1247 in
Table~\ref{tab1}.  It is an elliptical galaxy (Template 1.2) with no
detected extinction at a redshift of 0.88. The luminosity is $1.5
\times 10^{10}$ L$_{\sun}$ and it is an unlikely candidate for
HDF850.1.

WFPC2 3-593.0 and 3-593.2 correspond to source 1212 in
Table~\ref{tab1}.  This is a very late galaxy (Template 6.0) with an
E(B-V) of 0.4 at a redshift of 1.84.  It has a luminosity of $1.76
\times 10^{11}$ but the predicted 850 $\micron$ flux is only 0.16 mJy,
below the detected flux of 7 mJy.  This is a case where the available
parameter space of the analysis may have been inadequate to identify
this object as the SCUBA source.  The analysis picked the hottest SED
with a relatively high E(B-V).  If an even hotter SED were available it
might have given a better fit with a correspondingly higher E(B-V).
The combination of the hotter SED and higher E(B-V) could dramaticly
increase the expected 850 $\micron$ flux.  Of the two visible sources
3-593 (1212) is by far the better candidate for HDF850.1.

An alternative source is 1277 which lies closer than 1$\arcsec$ to the
position of HDF850.1 given by \citet{hug98}.  It is barely visible in
the F814W band but is a strong source in the F110W and F160W bands.  It
is a high redshift object (z = 4.8) with a extinction of E(B-V) = 0.3
and a very late template of 5.9.  Source 1277 has a luminosity of $1.49
\times 10^{11}$ L$_{\sun}$ and a predicted 850 $\micron$ flux of 0.125
mJy which is also below the observed flux.  The location of the VLA
source however makes this identification much less likely.

\citet{dun02} have proposed an identification of HDF850.1 as a source
located between 3-586.0 and 3-593.1 that is only visible after
subtraction of 3-586 in ground based K band images.  They also quote a
possible detection after subtraction in the F160W images used here.  In
this scenario the flux of the sub-mm source is lensed  with an
amplification factor of 3.  If this is the true source, it has been
missed in the analysis performed here and would only be visible through
special processing which is beyond the scope of this paper.

HDF850.4 and HDF850.5 lie very close to each other with a separation of
about 12$\arcsec$. The sources are in fact blended and must be
deconvolved \citep{ser02}.   The only nearby source with strong
calculated 850 $\micron$ flux is source 1108 which lies almost midway
between the sources.  Its calculated 850 $\micron$ flux is 1.36 mJy
close to the observed fluxes of 2.3 mJy and 2.1 mJy for HDF850.4 and
HDF850.5.  The source has a spectroscopic redshift of 0.19, an E(B-V)
of 0.6 and a luminosity of $1.55 \times 10^{11}$L$_\sun$.  If the shift
in position for HDF850.1 to lie on top of 3-593 is applied to HDF850.4
and HDF850.5, HDF850.5 would be within 4$\arcsec$ of 1108.  In this
case sources 896 and 905 could be identified with HDF850.4.  Their
combined predicted flux is equal to 0.89 mJy which is near to the
observed flux of 2.3 mJy for HDF850.4.

If the stated detection limit of 2 mJy is taken for the SCUBA
observations in the NHDF \citep{hug98} there is one source in
Table~\ref{tab1} that has a predicted 850 $\micron$ flux above the
limit but was not detected by the SCUBA observations. The source is the
ULIRG, source 800, discussed in \S~\ref{sss-ulirg}. It corresponds to
source 277.211 in TWS where its flux was predicted to be 1.5 mJy rather
the 3.46 mJy found here.  The present analysis found the same redshift,
E(B-V) and template as TWS, but the observed F110W and F160W magnitudes
were brighter.  If the ULIRG contains a AGN component, as is indicated
by the x-ray component,  the variation might be real but is of
surprising magnitude.

\subsubsubsection{VLA 3651+1221}

\begin{figure}
 
\plotone{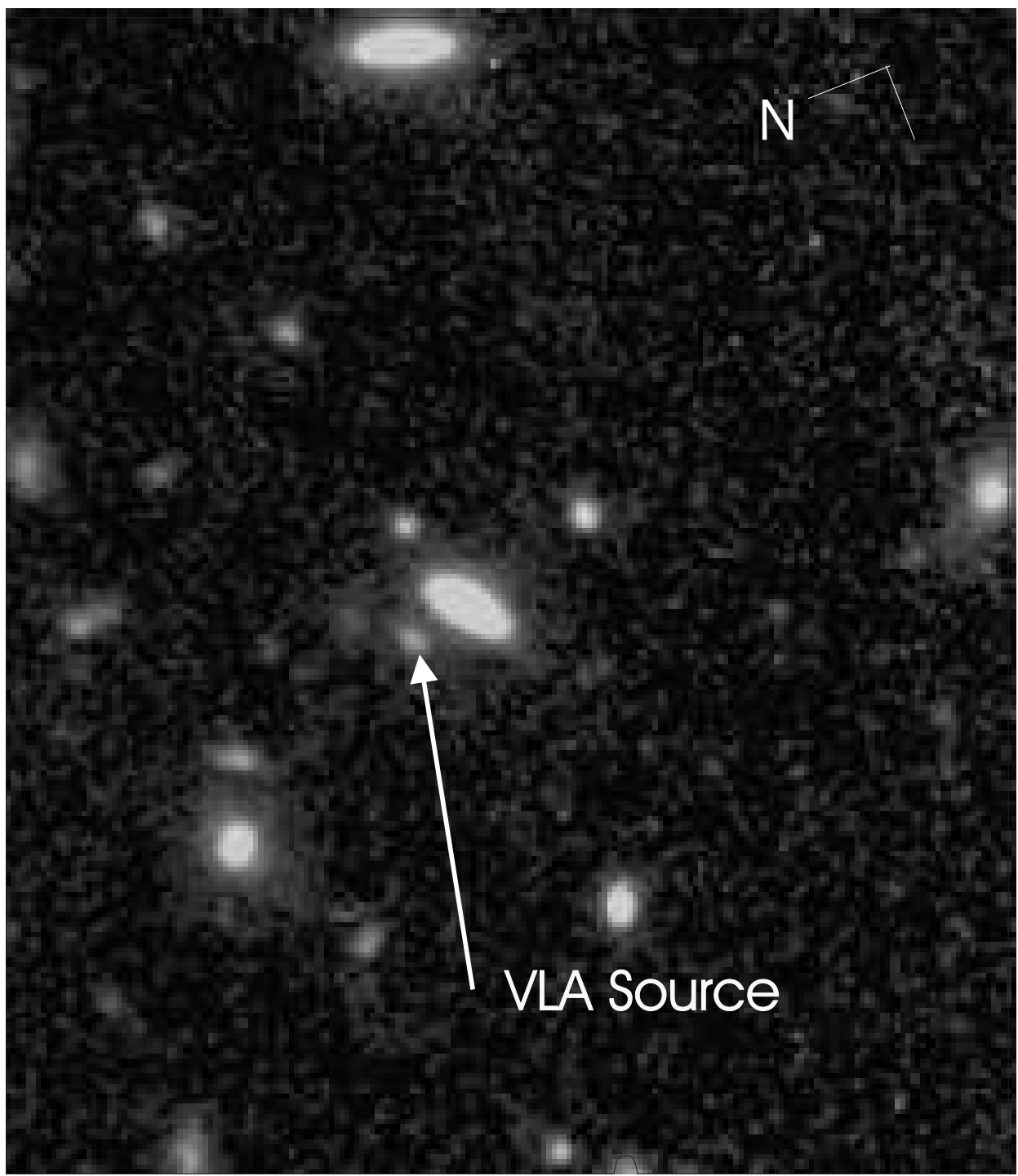}

\caption{The extremely red object associated with VLA 3651+1221.} 

\label{fig-vla}

\end{figure}

\citet{dwn99} observed a second source, VLA
3651+1221, very near the elliptical galaxy 3-659.1 which is source 1193
in Table~\ref{tab1}.  The F160W image shown in Figure~\ref{fig-vla}
clearly shows a strong source at the position of the VLA detection
that is very faint or absent in the F814W image.  It is clearly a
very red source.  The correspondence was first noticed by
\citet{dicet00} who claim that this is the second reddest source in the
NHDF. \citet{pap01} point out that the object is also an x-ray source
\citep{hor00, bra01a} and therefore probably an AGN. The SE source
extraction program linked the source with the elliptical galaxy in
producing Table~\ref{tab1}, which resulted in an identification of a
very late galaxy with a high extinction.  This is clearly an artifact
of the superposition of fluxes from two very different types of
galaxies.

\section{Concordance of Star Formation Rates} \label{s-con}

There has been significant debate \citep{bar00,bla99} on whether
optically based studies miss the majority of the star formation in the
universe because it is hidden in highly obscured galaxies.  The SFRs in
Figure~\ref{fig-sfr} for the optical, optical plus near infrared, and
submillimeter, however, are all in agreement.  On the face of it this
would seem to indicate that there is no large missing component of star
formation.  On the other hand the agreement may be fortuitous or the
definitions may be misleading. This section examines the various
studies to see what is missing and how the agreement has been
achieved.

The only inconsistent measurements in Figure~\ref{fig-sfr} are the
measurements of \citet{mad99} which show a fall off of star formation
at a redshift of 4 and generally lower SFR at all redshifts.  The SFRs
in \citet{mad99} are derived from the numbers presented in
\citet{mad98}, (note erroneous reference in \citet{mad99}) which do not
appear to be corrected for surface brightness dimming. The lack of
dimming correction appears to be the principal reason for the fall off
at high redshifts. The average extinction correction in \citet{mad99}
is A$_{1500}$ = 1.2, which is an underestimate according to this work
and accounts for the lower values at redshifts 1 to 2.

The extinction corrected SFRs at redshifts 3 and 4 from \citet{sti99},
corrected for the cosmology adopted here, are coincident with the SFRs
found in this work.  The extinction corrected values were found by
multiplying the uncorrected SFRs by 4.7, corresponding to an E(B-V) =
0.15 \citep{sti99}. The extinction correction value in \citet{sti99} is
well motivated and is interpreted there as the proper correction to the
UV flux from observed galaxies in the study.  It could also be
interpreted as a lower E(B-V) correction to the observed galaxies plus
a factor that accounts for galaxies not observed due to very high
extinction.  Without determining an extinction for each galaxy in the
sample it is not possible to separate the two interpretations.  It
still leaves the possibility that the actual average extinction to the
observed galaxies is less than E(B-V)= 0.15 and that there are other
undetected galaxies that contribute to the SFRs observed by
\citet{bar00}. Note that the comparison with the \citet{sti99} in
\citet{bar00} Figure 11 is to the uncorrected rate  and a $\Lambda = 0$
universe.

The SFRs in this work are both extinction corrected on an individual
galaxy basis and corrected for surface brightness dimming.  The
extinction correction at redshifts 2 and 4 are 4.9 and 6.1, near the
range of the \citet{sti99} correction. The surface brightness dimming
corrections are 2.0 in both cases, therefore, the extinction correction
is the dominant correction rather than galaxies or parts of galaxies
undetected due to surface brightness dimming. This would appear to
indicate that the observed sub-mm SFR can be accounted for by galaxies
that are observable in deep HST optical and near infrared observations
if not necessarily optical ground based observations.

It can also be argued that the present work has been unsuccessful in
identifying the 3 observed sub-mm sources in the fields of the 3 WF
chips. There is no doubt that there can be luminous galaxies that are
so highly extincted that even deep HST observations will miss
them.  At a redshift of 3 the F160W NICMOS band is measuring the 4000
$\AA$ flux which can be severely extincted.  It is also possible that
the optical depth that can be reached with the F160W filter is less
than the total optical depth of the galaxy so that the present analysis
determines a lower luminosity and extinction than is actually present
in the galaxy.  Another possibility is that the assumed dust SED is not
appropriate for  the bright sub-mm sources.  The luminosity in the 850
$\micron$ SCUBA band is a small fraction of the total dust luminosity, therefore,
small changes in the dust SED can greatly change the 850
$\micron$ luminosity without significantly affecting the total dust
luminosity.  Perhaps the bright sub-mm sources have colder dust SEDs
than the Arp 220 dust SED used here.  The bottom line is that on an
individual source level there are many uncertainties that can lead to
erroneous sub-mm predictions in the present analysis.  It should be
noted, however, that the sub-mm predictions are only intended to serve
as a check on the total sub-mm power not as a individual source
predictor.

In assessing the effect of underpredicting the bright sub-mm sources it
should be noted that the correction factor used in \citet{bar00} to
account for the fainter undetected sources is a factor of 11,therefore,
the bright sub-mm sources account for less than 10$\%$ of the SFR.
Based on the size of the correction it might be argued that the
majority of star formation is also missed by the sub-mm observations.
The correction to the sub-mm observations depends on the assumed
luminosity function whereas the correction to the optical and near IR
observations depends on the assumed extinction law.  Even if all of
the observed sub-mm sources are missed by optical and near-IR studies,
the error in the SFR is small.  It can be argued that at the depth of
the HST observations only the most obscured sources are missed and that
the bulk of the galaxies that produce the sub-mm emission and the
majority of the SFR are detected and on the average properly accounted
for.  Within the caveats  discussed above it appears that although some
of the extreme sub-mm sources are missed by the optical and near
infrared observations, the total SFR derived from optical, optical and
near infrared, and sub-mm observations are in agreement indicating that
the majority of the SFR is accounted for in the three types of
observations.  All of these methods however, have correction factors on
the order of a factor of 10 based on assumptions that still need to be
confirmed.

\section{Summary} \label{sec-sum}

Analysis of HST archival data from WFPC2 and NICMOS in the NHDF
yields photometric redshifts, extinctions and SEDs for almost 2000
galaxies.  From this data the extinction corrected SFR for each galaxy
was determined. After correction for surface brightness dimming and the
variable sensitivity over the NICMOS detector arrays, the star
formation history of the NHDF from a redshift 1 to 6 shows a roughly
constant SFR.  Optical studies of Lyman break galaxies and sub-mm
observations produce SFRs that are consistent with the results in this
study.  The sub-mm background predicted from our analysis is consistent
with the observed background, indicating that the analysis has not
missed a significant star forming component due to high extinction.
The inability to accurately predict the fluxes and locations of the
observed NHDF sub-mm sources does indicate that some bright sub-mm
sources are certainly missed by this work but that these are not the
sources that contribute the majority of star formation in the field.

\section{Acknowledgements}

The author would like to acknowledge the significant contributions to
this work by Ray Weymann and Lisa Storrie-Lombardi in the initial data
reduction and analysis as well as their contributions in establishing
the methodology which was carried over from TWS.  The author would also
like to thank Mark Dickinson and all of his group who planned and
executed the NICMOS observations of the entire Northern HDF in a
General Observer program. This work is supported in part by NASA grant
NAG 5-10843.  This article is based on observations with the NASA/ESA
Hubble Space Telescope, obtained at the Space Telescope Science
Institute, which is operated by the Association of Universities for
Research in Astronomy under NASA contract NAS5-26555.

\clearpage

\begin{deluxetable}{ccccccccc}
\tablecaption{Galaxies with Redshift Errors Greater than 0.5. 
\label{tab-err}}
\tablewidth{0pt}
\scriptsize
\tablehead{\colhead{ID} & \colhead{WFPC-ID} & \colhead{Phot. Z} & 
\colhead{Spec. z}  & \colhead{Template} & \colhead{E(B-V)} &
\colhead{Qual. \tablenotemark{a}} & \colhead{Sp. \tablenotemark{b}} &
 \colhead{fail \tablenotemark{c}}}
\startdata
6 & 4-916 & 0.08 & 0.904 & 6.0 & 0.6 & 9 & A & P \\
13 & 4-928 & 0.48 & 1.015 & 6.0 & 1.0 & 4 & E & P \\
84 & 4-878 & 0.0 & 0.892 & 3.0 & 0.04 & 4 & E & S,P \\
108 & \nodata & 1.84 & 0.584 & 5.2 & 0.5 & 3 & A & S \\
339 & 4-445 & 1.84 & 2.5 & 5.8 & 0.3 & 2 & EA & S \\
369 & \nodata & 1.92 & 2.801 & 5.9 & 0.3 & 3 & A & \nodata \\
524 & 2-251 & 0.24 & 0.962 & 4.3 & 0.4 & 6 & Q & \nodata \\
625 & 2-201 & 1.84 & 1.313 & 5.0 & 0 & 1 & E & S,P \\
1226 & 2-531 & 0.56 & 1.087 & 5.8 & 0.5 & 3 & A & F,P \\
1656 & 2-982 & 0.56 & 1.147 & 5.7 & 0.4 & 1 & E & F \\
1762 & \nodata & 1.12 & 0.47 & 3.5 & 0.0 & 1 & E & P \\ 
\tablenotetext{a}{This is the quality of the spectrum as noted in 
\citet{chn00} 
where 1 is the highest 11 is the lowest cited in table 2b of that work.}
\tablenotetext{b}{Sp is the spectral type listed in \citet{chn00}. E is 
an 
emission line galaxy, A is an absorption line galaxy, EA is a galaxy 
displaying 
both emission and absorption and Q is a broad absorption line galaxy}
\tablenotetext{c}{Fail is the probable cause of failure. P is hitting a 
boundary 
in parameter space, S is the probable superposition of two galaxies, and 
F is 
the problem associated with z = 0.56 discussed in the text.}
\enddata
\end{deluxetable}
\clearpage

\begin{deluxetable}{cccccccccccccccc}
\rotate
\tabletypesize{\tiny}
\tablecaption{Listing of measured quantities \label{tab1}}
\tablewidth{0pt}
\tablehead{
\colhead{NICMOS} & \colhead{WFPC} & \colhead{z} & \colhead{E(B-V)} & 
\colhead{SFR} &  \colhead{Lum.} & \colhead{frac.\tablenotemark{a}}  & 
\colhead{ISO6} & \colhead{ISO15} & \colhead{SCUBA}  &  \colhead{T 
\tablenotemark{b}} & \colhead{$\chi^2$} & 
\colhead{Tot.\tablenotemark{c}} 
& 
\colhead{Ap.\tablenotemark{d}} & \colhead{RA} & \colhead{DEC} \\ 
\colhead{ID} & 
\colhead{ID} & \colhead{} & \colhead{} & \colhead{M$_\odot$} & 
\colhead{L$_\odot$} & \colhead{} & \colhead{6 \micron} & \colhead{15 
\micron} 
& 
\colhead{850 \micron} & \colhead{} & \colhead{ }& \colhead{mag} & 
\colhead{mag} 
& \colhead{12\fh} & \colhead{+62\fdg} \\ 
\colhead{} & \colhead{} & \colhead{} & 
\colhead{} & \colhead{yr$^{-1}$} & \colhead{} & \colhead{} & 
\colhead{flux} & 
\colhead{flux} & \colhead{flux} & \colhead{} & \colhead{} & \colhead{} & 
\colhead{} & \colhead{36\fm} & \colhead{} \\ \colhead{} & \colhead{} & 
\colhead{} & \colhead{} & \colhead{} & \colhead{} & \colhead{} & 
\colhead{mJy} 
& 
\colhead{mJy} & \colhead{mJy} }
\startdata
    1 &4-951.0& 0.00& 0.50& 2.831 &0.00E+00& 0.86 &0.00E+00 &0.00E+00 &0.00E+00& 5.6& 3.0  & 26.4  & 27.4 &38.23&12:28.3 \\
   2 &\nodata& 2.88& 0.00& 0.258 &3.64E+09& 0.00 &0.00E+00 &0.00E+00 &0.00E+00& 3.5& 18.  & 28.2  & 27.2 &38.24&12:31.7 \\
   3 &\nodata& 2.56& 0.00& 0.131 &2.15E+09& 0.00 &0.00E+00 &0.00E+00 &0.00E+00& 3.3& 19.  & 28.2  & 27.3 &38.32&12:34.2 \\
   4 &\nodata& 2.56& 0.10& 0.691 &3.94E+09& 0.39 &5.16E-06 &7.82E-06 &1.61E-03& 5.1& 5.6  & 28.7  & 27.2 &38.32&12:32.2 \\
   5 &\nodata& 2.80& 0.00& 1.984 &7.36E+09& 0.00 &0.00E+00 &0.00E+00 &0.00E+00& 5.8& 19.  & 27.1  & 27.1 &38.49&12:35.5 \\
   6 &4-916.0& 0.90& 0.60& 0.182 &6.34E+08& 0.95 &1.90E-05 &1.61E-04 &8.75E-04& 6.0& 11.  & 22.5  & 24.6 &38.59&12:33.8 \\
   7 &4-922.0& 3.20& 0.20& 2.937 &1.18E+10& 0.53 &0.00E+00 &2.10E-05 &6.76E-03& 5.6& 2.6  & 27.8  & 27.7 &38.60&12:29.1 \\
   8 &4-952.0& 3.52& 0.20& 5.092 &2.15E+10& 0.51 &0.00E+00 &3.00E-05 &1.29E-02& 5.5& 4.6  & 27.4  & 27.3 &38.77&12:18.8 \\
   9 &4-930.0& 0.48& 0.30& 0.662 &2.46E+09& 0.75 &2.63E-04 &1.76E-03 &4.70E-03& 5.8& 8.1  & 24.2  & 25.7 &38.79&12:25.9 \\
  10 &4-953.0& 2.88& 0.00& 0.587 &2.62E+09& 0.00 &0.00E+00 &0.00E+00 &0.00E+00& 5.4& 2.2  & 27.5  & 27.9 &38.83&12:17.9 \\
\enddata
\tablenotetext{a}{This is the fraction of the luminosity removed by 
extinction 
and re-emitted in the mid and far infrared}
\tablenotetext{b}{The selected template number between 1.0 (early-cold) 
and 
6.0 
(late-hot).}
\tablenotetext{c}{The total AB magnitude in the F160W filter.}
\tablenotetext{d}{The F160W magnitude in an $0.6\arcsec$ diameter 
aperture.}
\end{deluxetable}

The remainder of the 1927 table entries are available in the electronic
version of the Astrophysical Journal publication.

\clearpage
\begin{deluxetable}{ccccccc}
\tablecaption{Properties of galaxies with F160W magnitudes significantly 
brighter than predicted L* magnitudes. \label{tab-lstar}}
\tablewidth{0pt}
\scriptsize
\tablehead{\colhead{ID} & \colhead{WFPC-ID} & \colhead{Redshift} & 
\colhead{F160W AB mag.} & \colhead{E(B-V)} & \colhead{Template} &
\colhead{Lum.}}
\startdata
1636.0 & 3-839.0 & 5.2 & 23.8 & 0.02 & 5.1 & 5.72E11 \\
398.0 & 4-555.11 & 3.04 & 24.4 & 0.08 & 5.6 & 2.7E11 \\
410.0 & 4-555.1 & 2.799 & 23.8 & 0.20 & 5.7 & 6.09E11 \\
81.0 & 4-713 & 4.08 & 23.6 & 0.00 & 3.9 & 4.96E11 \\
1604.0 & 3-82 & 4.32 & 24.3 & 0.10 & 3.6 & 6.5E11 \\
1630.0 & 3-367 & 4.48 & 24.8 & 0.06 & 3.5 & 3.66E11 \\
1409.0 & 2-591 & 2.96 & 24.4 & 0.00 & 5.2 & 8.84E10\\
1179.0 & 2-578 & 3.92 & 25.0 & 0.00 & 5.2 & 1.09E11 \\
812.0 & 4-169 & 5.04 & 25.2 & 0.10 & 3.9 & 3.38E11 \\
\enddata
\end{deluxetable}
\clearpage

\begin{deluxetable}{ccccc}
\tablecaption{SFR corrections for extinction and surface brightness dimming 
\label{tab-sfr}}
\tablewidth{0pt}
\scriptsize
\tablehead{\colhead{z} & \colhead{Uncor.} & \colhead{Ext. Cor.} &
\colhead{Ext.\& Dim.} & \colhead{\% Lum. \tablenotemark{a}} \\
\colhead{} & \colhead{SFR} & \colhead{SFR} & \colhead{Cor. SFR}
&\colhead{Missing}}

\startdata
1 & 6.8E-3 & 2.8E-1 & 2.8E-1 & 0\% \\
2 & 3.7E-2 & 1.8E-1 & 3.7E-1 & 50\% \\
3 & 5.0E-2 & 8.9E-2 & 9.6E-2 & 8\% \\
4 & 1.0E-1 & 6.1E-1 & 1.2E-1 & 50\% \\
5 & 1.8E-2 & 6.5E-2 & 2.3E-1 & 72\% \\
6 & 3.5E-2 & 8.1E-2 & 3.0E-1 & 73\% \\
\enddata
\tablenotetext{a}{This is the percentage of missed luminosity between 
the 
extinction corrected SFR and the extinction corrected SFR also corrected 
for 
surface brightness dimming.}
\end{deluxetable}
\clearpage

\begin{deluxetable}{ccccccc}
\footnotesize
\tablecaption{Sources of Numerical Variances 
\label{tab-ers}}
\tablewidth{0pt}
\tablehead{
\colhead{Statistic} & \colhead{z = 1} & \colhead{z = 2} & \colhead{z = 
3} 
& 
\colhead{z = 4} & \colhead{z = 5} & \colhead{z = 6}}
\startdata
Number of Sources & 495 & 370 & 339 & 160 & 45 & 55 \\
$1/\sqrt{N}$ & 0.045 & 0.052 & 0.058 & 0.079 & 0.15 & 0.13 \\
I$_2$ & 0.16 & 0.11 & 0.13 & 0.16 & 0.19 & 0.21 \\
$\sigma_N/N$ & 0.09 & 0.11 & 0.13 & 0.16 & 0.19 & 0.21 \\
$16 \%$ Confidence fraction & 0.47 & 0.48 & 0.21 & 0.27 & 0.64 & 0.82 \\
$84 \%$ Confidence fraction &0.37 & 0.21 & 0.69 & 1.2 & 1.1 & 0.13 \\ 
\enddata 
\end{deluxetable}

\end{document}